\documentclass[fleqn,usenatbib]{mnras}

\usepackage{newtxtext,newtxmath}
\usepackage[T1]{fontenc}

\DeclareRobustCommand{\VAN}[3]{#2}
\let\VANthebibliography\thebibliography
\def\thebibliography{\DeclareRobustCommand{\VAN}[3]{##3}\VANthebibliography}

\usepackage{graphicx}	% Including figure files
\usepackage{amsmath}	% Advanced maths commands
\graphicspath{{./}{figures/}}

\usepackage{xcolor}

\newcommand{\ud}{\mathrm{d}}
\newcommand{\msunh}{h^{-1}\mathrm{M}_\odot }

\title[depletion halo model]{A physical and concise halo model based on the depletion radius}

\author[Zhou \& Han]{
Yifeng Zhou$^{1,2,3}$%\thanks{yifengzhou@sjtu.edu.cn} 
and Jiaxin Han$^{1,2,3}$\thanks{jiaxin.han@sjtu.edu.cn}
\\
$^{1}$ Department of Astronomy, School of Physics and Astronomy, Shanghai Jiao Tong University, Shanghai 200240, China\\
$^{2}$ Key Laboratory for Particle Astrophysics and Cosmology (MOE), Shanghai 200240, China\\
$^{3}$ Shanghai Key Laboratory for Particle Physics and Cosmology, Shanghai 200240, China\\
}

\date{Accepted XXX. Received YYY; in original form ZZZ}

\pubyear{2023}

\begin{document}
\label{firstpage}
\pagerange{\pageref{firstpage}--\pageref{lastpage}}
\maketitle

\begin{abstract}
We develop a self-consistent and accurate halo model by partitioning matter according to the depletion radii of haloes. Unlike conventional models that define haloes with the virial radius while relying on a separate exclusion radius or ad-hoc fixes to account for halo exclusion, our model distributes mass across all scales self-consistently and accounts for both the virialized and non-virialized matter distribution around each halo. Using a cosmological simulation, we show that our halo definition leads to very simple and intuitive model components, with the one-halo term given by the Einasto profile with no truncation needed, and the halo-halo correlation function following a universal power-law form down to the halo boundary. The universal halo-halo correlation also allows us to easily model the distribution of unresolved haloes as well as diffuse matter. Convolving the halo profile with the halo-halo correlation function, we obtain a complete description of the halo-matter correlation across all scales, which self-consistently accounts for halo exclusion at the transition scale. Mass conservation is explicitly maintained in our model, and the scale dependence of the classical halo bias is easily reproduced. Our model can successfully reconstruct the halo-matter correlation function within an accuracy of $9\%$ for halo virial masses in the range of $10^{11.5}h^{-1}{\rm M}_{\odot}<M_{\rm vir}<10^{15.35}h^{-1}{\rm M}_{\odot}$ at $z=0$, and covers the radial range of $0.01h^{-1}{\rm Mpc}<r<20h^{-1}{\rm Mpc}$. We also show that our model profile can accurately predict the characteristic depletion radius at the minimum bias and the splash-back radius at the steepest density slope locations.
\end{abstract}

\begin{keywords}
 large-scale structure of Universe -- dark matter -- galaxies: haloes
\end{keywords}

%%%%%%%%%%%%%%%%%%%%%%%%%%%%%%%%%%%%%%%%%%%%%%%%%%

%%%%%%%%%%%%%%%%% BODY OF PAPER %%%%%%%%%%%%%%%%%%

\section{Introduction}

%\jx{many of the references do not have hyperlinks. Please use bibtex record downloaded from ADS to fix this.}

% \jx{suggested outline for intro: 

% - introduce halo model: powerful and important. works well on both small and large scale. many applications (cosmology, galaxy-halo connection, clustering/lensing profile ...)

% - issue: transition scale ambiguously described, with ad-hoc fixes. for example: Hayashi$\&$White; Zentner; vdB.. and mass conservation is not necessarily satisfied?

% - This is ultimately linked to the ambiguity of the halo boundary: physical and geometrical considerations. many recent developments: various boundaries; Garcia;

% - depletion boundary: physical and geometrically relevant. We base our approach on this. Also build the model fully self-consistently from first principle/data-driven recipes, by rebuilding the one halo profile, halo-halo corr, halo bias, mass func accordingly.

% the original intro is concrete but too involved, and somewhat out of focus. This work is not to introduce a new boundary, but to demonstrate that the depletion radius can solve the halo model problem, and lead to a much more concise halo model.
% }

The halo model of the large scale structure \citep[see][for a review]{cooray2002halo} is a powerful analytic framework for describing the distribution of dark matter in the Universe. At the largest scales, the distribution of dark matter carries information about the initial density field and history of the universe. At galactic scales, dark matter dominates the potential of the virialized matter, thus determining the formation and evolution of galaxies. By assuming the mass in the universe are all bounded into individual dark matter haloes, the halo model successfully describes the matter distribution on both small and large scales. %In this model framework, the matter clustering signal is decomposed into two parts: the 1-halo term originating from the internal structure of haloes on small scale, and the 2-halo term originating from the clustering of neighbouring haloes on large scale. 
It has been widely used in studying the matter-matter power spectrum \citep[e.g.,][]{2003MNRAS.341.1311S, 2012ApJ...761..152T}, the galaxy-halo connection \citep[e.g.,][]{2005MNRAS.362.1451M,2012MNRAS.426..566C}, and the galaxy-galaxy correlation \citep[e.g.,][]{2000MNRAS.318.1144P,2000MNRAS.318..203S}. 
%The distribution of dark matter is a critical question for the $\Lambda$CDM cosmology. At the largest scales, the distribution of dark matter carries information about initial density field of the universe and provides a test for the linear theory; at galactic scales, dark matter dominants the dynamics around potential well, thus determines the formation and evolution of galaxies. Based on the halo model \citep{cooray2002halo}, the matter density field can be built up by the spherically symmetric dark matter haloes, so that it is essential to construct an accurate description of spherically averaged density profile. Using the spherical collapse top-hat model, \citet{gunn1972infall} made a simple prediction of the density profile of collapsed haloes. With in-depth studies about dark matter haloes in simulations, some analytical fitting functions have been proposed to describe the density profile inside the virial radius,  such as Navarro–Frenk–White \citep[][hereafter, NFW]{navarro1995simulations,navarro1996structure,navarro1997universal} formula or Einasto \citep{einasto1965construction,einasto1969andromeda} formula.

Despite its success at small and large scales, the halo model has encountered difficulties in accurately describing the matter distribution on the intermediate scale, that is, around the boundary of haloes. By construction, haloes are independent objects that do not overlap with each other, an effect known as halo exclusion. However, practical definitions of haloes may not guarantee that they can fully partition the entire density field. As a result, ambiguities in describing the matter distribution around the halo boundary arise, and different implementations of the model usually have to introduce ad-hoc fixes to improve the model accuracy around this scale. % depending on how the density profile of a halo is treated outside its boundary, and how neighbouring haloes are modelled around the halo edges.
For example, \citet{hayashi2008understanding} splices the density profile where the values of 1-halo and 2-halo terms are equal. \citet{tinker2005mass} introduced a distance dependent exclusion probability in the calculation of the matter-matter correlation function, which is motivated by the ellipsoidal shapes of haloes. \citet{van2013cosmological}, on the other hand, introduced a parametric radial-dependent halo bias and a truncation radius to fit the two-point correlation function. %\yf{Particularly, the latter two works make additional calibrations (e.g., non-sphericity and alignment of haloes) in the 2-halo term, which may lead to inconsistencies of the halo profile in the 1-halo and 2-halo terms.}

%Despite these efforts to improve the model accuracy, none of them bring a physical view of the exclusion radius. 
The ambiguity of the halo boundary also leads to other global issues in the halo model. For example, it can be challenging to explicitly impose mass conservation in the halo model. Some recent works \citep{2016PhRvD..93f3512S, 2020A&A...641A.130M,2020PhRvD.101j3522C,2021MNRAS.502.1401M} have noticed this problem, but the physical meaning of their solutions are yet to be fully understood. Besides, to compensate at least partly for the ambiguity of the halo size, additional fitting functions and phenomenological parameters have to be introduced in many components of the halo model~\citep[e.g.,][]{UniversalMF,nonUniversalMF, concentrationDimer}% \yf{Can you check if the references here are appropriate or need to add more?}\jx{first two are good but not sure about the rest}
, which further complicates the model.

These complications are ultimately caused by the lack of a matching halo boundary in the halo model. In particular, the classical virial definition of a halo radius provides a useful reference for identifying the virialized part of a halo~\citep{gunn1972infall}, but is not optimized for partitioning the space with haloes~\citep{garcia2022better}. In fact, the virial radius itself faces several theoretical difficulties in serving as a physical halo radius. For example, it has been shown that this radius may not correctly demarcate the virialized part of a halo~\citep{2014ApJ...792..124Z, StaticRadius}, and could introduce unphysical (or pseudo) evolution of halo properties \citep{diemer2013pseudo}. More importantly, such an approximate description crudely separates the halo from the background Universe at the virial radius and ignores any transition region towards its non-virialized environment~\citep{fong2021natural}, nor does it account for the dynamic growth of a halo. For example, tidal stripping of satellite haloes can happen well outside the virial radius of the host halo \citep{bahe2013does,behroozi2014mergers}, and ejected subhaloes can be found a few times outside the virial radius of the host halo~\citep{2009ApJ...692..931L}. In addition,  the aspherical shape of haloes \citep{jing2002triaxial,allgood2006shape,mansfield2017splashback,wang2022anisotropy} means the spherically averaged density profile of a halo does not stop abruptly at the virial radius. %The ambiguities of the halo definition cause the complication at intermediate scales where the 1-halo and 2-halo terms mix. 

To clarify the matter distribution around the halo boundary and to find a more physical and intrinsic characterisation of halo size, a number of recent works have been devoted to finding new definitions of the halo radius. % out of physical and geometrical considerations. 
Considering that a halo is a growing system, the radius when an infalling particle first reaches the apocentre of its orbit, the so-called splashback radius~\citep{1984ApJ...281....1F, 1985ApJS...58...39B}, has been revived to provide a more physical and extended characterisation of the halo boundary \citep{diemer2014dependence,adhikari2014splashback,more2015splashback,shi2016outer,mansfield2017splashback}. Such a radius is typically found at the steepest density slope location, and depends on both the halo mass and the mass accretion rate \citep{more2015splashback,diemer2017splashback,contigiani2021mass,o2021splashback}. \citet{diemer2020universal} found that the halo mass functions are significantly more universal when haloes are defined using the splashback radius. Along the same line, many recent developments have moved their attention to the dynamical structure around haloes. \citet{aung2021phase} defined an ``edge radius'' according to how well the infalling and orbiting substructures around a halo can be separated in phase space. Most recently, \citet{diemer2021dynamics} has developed a robust algorithm for separating particles into infalling and orbiting components according to the pericentric passage of each particle. A similar decomposition was obtained from \citet{garcia2022better} by additionally considering the accretion time of particles. They argued that a physical halo collects all particles orbiting in their self-generated potential. 

Focusing on the evolution of the density profile around the halo, \citet{fong2021natural} proposed an independent characterisation of the halo size named depletion radius. As a halo accretes matter from its neighbourhood, its own density grows in the outer part, while the density of the surrounding environment gets depleted over time. More specifically, the transition between the growing and decaying part of the density profile happens exactly where the mass infall rate is the maximum, which is defined as the ``inner depletion radius", $R_{\rm id}$. This radius can also be found near the minimum in the halo bias profile, which leads to the definition of a twin radius called characteristic depletion radius at the bias minimum, $R_{\rm cd}$. Recent works have measured $R_{\rm id}$ and $R_{\rm cd}$ from observations. \citet{Fong2022First} made the first measurements of $R_{\rm cd}$ using weak lensing observations. \citet{li2021outermost} measured the $R_{\rm id}$ of the Milky Way using the motion of nearby dwarf galaxies. %These observations made it necessary to develop a halo model including the depletion radius.

\citet{fong2021natural} argued that the bias minimum can be interpreted both as a consequence of the depletion process, and as a manifestation of halo exclusion. As a result, the depletion radius may potentially serve as the desired exclusion radius in the halo model. In fact, by solving for an optimal exclusion radius with a flexible halo model, \citet{garcia2021redefinition} found a halo radius that happens to be very close to the inner depletion radius of \citep{fong2021natural}. However, the \citet{garcia2021redefinition} model involves some assumptions and approximations that may not match the exact behavior of the depletion radius. It thus remains to be seen how well the depletion radius defined physically in the first place performs in the halo model.

This paper intends to demonstrate that the depletion radius can indeed solve the halo exclusion problem, and lead to a much more concise halo model. By introducing the inner depletion radius as the natural boundary of haloes, we build the halo model fully self-consistently from first principles. It is achieved by rebuilding the one-halo profile, halo-halo correlation functions, halo bias, and halo mass function accordingly. We construct a halo catalog consistent with the new definition to investigate their statistical properties. We use the resulting halo model to reconstruct the spherical distribution of matter around haloes across scales and try to extract information about characteristic radii from the predicted profiles.

This paper is organized as follows. In Section \ref{sec:sec2}, we introduce the simulation data and the approach to measure the inner depletion radius. Based on the radius relation obtained from stacked velocity profiles, we construct an isolated halo sample by removing the haloes overlapping with more massive neighbours. In Section \ref{sec:sec3}, we describe the analytical framework of a halo model based on the depletion boundary. In this section, we investigate the halo-halo correlation function, the halo abundance, and the halo density profile in our halo catalog. Section \ref{sec:sec4} presents the fits to bias profiles.% of our model \jx{what do you mean: and features of a bias profile including the 2-halo term. }
We discuss a few open questions and complications about the model in Section \ref{sec:sec5}. %one concerns optional exclusion halo-halo radius and another focus on the characteristic depletion radius.
Finally, in Section \ref{sec:sec6}, we summarize the results of this paper.

\section{Simulation and halo samples}
\label{sec:sec2}
\subsection{Simulation data}
We use a $\Lambda$ cold dark matter simulation from the CosmicGrowth~\citep{jing2019cosmicgrowth} simulation suite with cosmological parameters $\Omega_{\rm m}=0.268$ and  $\Omega_{\Lambda}=0.732$. The simulation was run in a box of 600 $h^{-1}{\rm Mpc}$ per side containing $3072^3$ particles. Candidate haloes are identified with the Friends-of-friends (FoF) algorithm with a standard linking parameter of $0.2$, and processed with ${\rm HBT}+$ \citep{han2012resolving,han2018hbt+} to obtain subhaloes. 

We collect host haloes with at least $10^{11.5} h^{-1}{\rm M}_{\odot}$ (or $\sim 500$ particles) inside the virial radius at $z=0$, resulting in a fiducial halo sample of $2\times 10^6$ haloes. These haloes are further divided into seven mass bins equally spaced in logarithmic mass, labelled with MB1 through MB7 as detailed in Table~\ref{tab:halopopulation}, covering a mass range of $11.5<{\rm log}_{10}[M/(h^{-1}{\rm M}_{\odot})]<15.35$. To build a halo model based on the depletion radius self-consistently, these haloes are further processed to remove overlapping ones on the depletion scale. Such a cleaning process will be discussed in more detail in Section~\ref{sec:exclusion}.

We use the microscopic bias profile to describe the mass distribution around a halo centre. According to \citet{han2019multidimensional}, the microscopic bias profile around an individual halo is defined as
\begin{equation}
    \beta(r) = \frac{\delta(r)}{\xi_{\rm mm}(r)}
    \label{eq:mic_bias}
\end{equation}
where $\delta(r)\equiv\rho(r)/\rho_{\rm m}-1$ is the overdensity profile of matter around each halo, $\rho_{\rm m}$ is the mean matter density of the universe. $\xi_{\rm mm}(r)$ is the matter-matter correlation function. Stacking the microscopic bias profile in a given halo bin, we obtain the average bias profile
\begin{equation}
    b(r) = \langle \beta(r)\rangle = \frac{\langle \delta(r)\rangle}{\xi_{\rm mm}(r)} = \frac{\xi_{\rm hm}(r)}{\xi_{\rm mm}(r)}
    \label{eq:bias_prof}
\end{equation}
where $\xi_{\rm hm}(r)$ is the halo-matter correlation function, and $\langle \dots\rangle$ represents the averaging over all the haloes in each halo mass bin. Using bias profiles rather than density profiles or overdensity profiles to describe the matter distribution facilitates us to highlight some important features in the halo model, such as the linear halo bias and the characteristic depletion radius. 

\begin{figure}
    \includegraphics[width=\columnwidth]{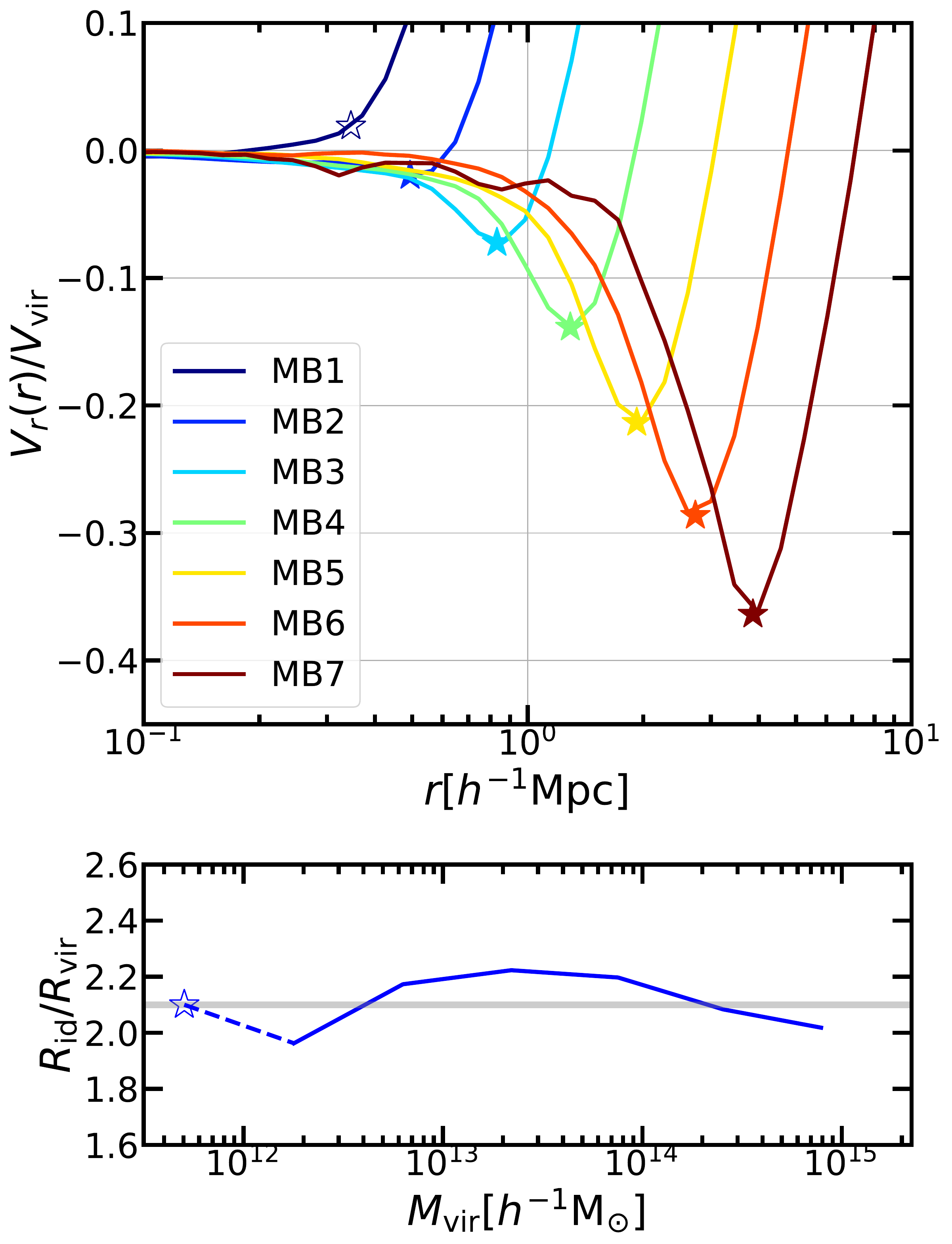}
    \caption{The inner depletion radius of different mass bins. \textit{Top}: the velocity profile of each mass bin. Stars mark positions of the minimum of the profiles, namely the inner depletion radius $R_{\rm id}$. For MB1 which does not show a minimum, $R_{\rm id}$ is estimated from $2.1R_{\rm vir}$. \textit{Bottom}: the ratio between $R_{\rm id}$ and the median $R_{\rm vir}$ in each mass bin (blue solid and dashed line). The gray solid line marks the ratio of $2.1$. The star marks the $R_{\rm id}$ in MB1 estimated according to this ratio.}
    \label{fig:r_ratio}
\end{figure}

The inner depletion radius, $R_{\rm id}$, is defined as the location of the maximum mass inflow rate,  which is nearly identical to the location of the minimum of the radial velocity profile. The total radial velocity can be written as $v_{\rm r}=v_{\rm p}+v_{\rm H}$, where $v_{\rm p}$ and $v_{\rm H}$ are the peculiar and Hubble velocity respectively. The average radial velocity profile for each of our halo mass bins is shown in the top panel of Figure \ref{fig:r_ratio}. The profiles all show a trough of negative radial velocity, corresponding to the region of infalling material, except for the lowest mass bin, MB1. For MB1, the average radial velocities of particles monotonically increase with radius with no infall region, leading to the difficulty in defining its inner depletion radius. Fortunately, the ratio of $R_{\rm id}$ and $R_{\rm vir}$ is approximately a constant of $2.1$ for six other mass bins, as shown in the bottom panel of Figure \ref{fig:r_ratio}, in agreement with \citet{fong2021natural}. We thus use $2.1R_{\rm vir}$ as the proxy of $R_{\rm id}$ for MB1. The relation of $R_{\rm id}\simeq 2.1R_{\rm vir}$ is also useful for estimating the inner depletion radius of individual haloes, as the significant noise makes it hard to determine the minimum of the individual velocity profile. For the remaining six mass bins, we implement a local polynomial interpolation to find the minimum velocities and define $R_{\rm id}$s by using the minimum data point and its two adjacent points.

%\subsection{Numerical methods}
%\rev{We determine the Characteristic/inner depletion radius by searching for the minimum of the bias/velocity profile. We perform a second-order polynomial interpolation to find the minimum and its location using the minimum data point and its adjacent two points. Then $R_{\rm cd }$/$R_{\rm id}$ is determined as the position of the minimum value of the polynomial interpolated from the bias/velocity profile.}

%\rev{In this paper, there are several places that need to determine the unknown parameters. Here we give a quick description of our fitting methods. we perform functional fits to obtain the unknown parameters. To obtain a fit of equal quality at the whole range we considered, we construct the merit function as the sum of the square logarithmic differences for all fits, minimized using a Python package \texttt{iminuit}\footnote{\url{https://doi.org/10.5281/zenodo.3949207}}. We specify the input data and details in the following where the fits are performed.}

\subsection{Exclusion criterion and halo samples}
\label{sec:exclusion}
When extending the boundary of the dark matter halo to $R_{\rm id}$, the halo catalog generated by the FoF algorithm needs to be cleaned for overlapping haloes according to this new boundary. Such a cleaning process is not only a mathematical requirement, but also out of physical considerations. As discussed in \citet{fong2021natural} and \citet{Gao23Depletion}, the active accretion region of a typical halo is created by its own gravity. If two haloes are too close so that their depletion regions overlap, the accretion features of the smaller ones can be obscured and even destroyed by the massive neighbour. In this case, it is more consistent to define the small halo as a subhalo and exclude it from the halo sample. %Such a physical interpretation is identical to the exclusion effect of haloes, i.e. if the separation of two haloes is $d<r_1+r_{\rm decay}$, the smaller halo will be treated as a subhalo. T
Specifically, we use $R_{\rm id}=2.1R_{\rm vir}$ to estimate the inner depletion radius of each halo, and exclude it whenever $d<R_{\rm id}+R_{\rm id,ngbr}$ where $R_{\rm id,ngbr}$ is the radius of a more massive neighbouring halo and $d$ is the distance to it. This is the so-called hard-sphere exclusion scheme \citep{garcia2019halo}. % In practice,  of the individual halo to avoid analyzing the velocity profiles with significant noise.

\begin{figure*}
    \includegraphics[width=\textwidth]{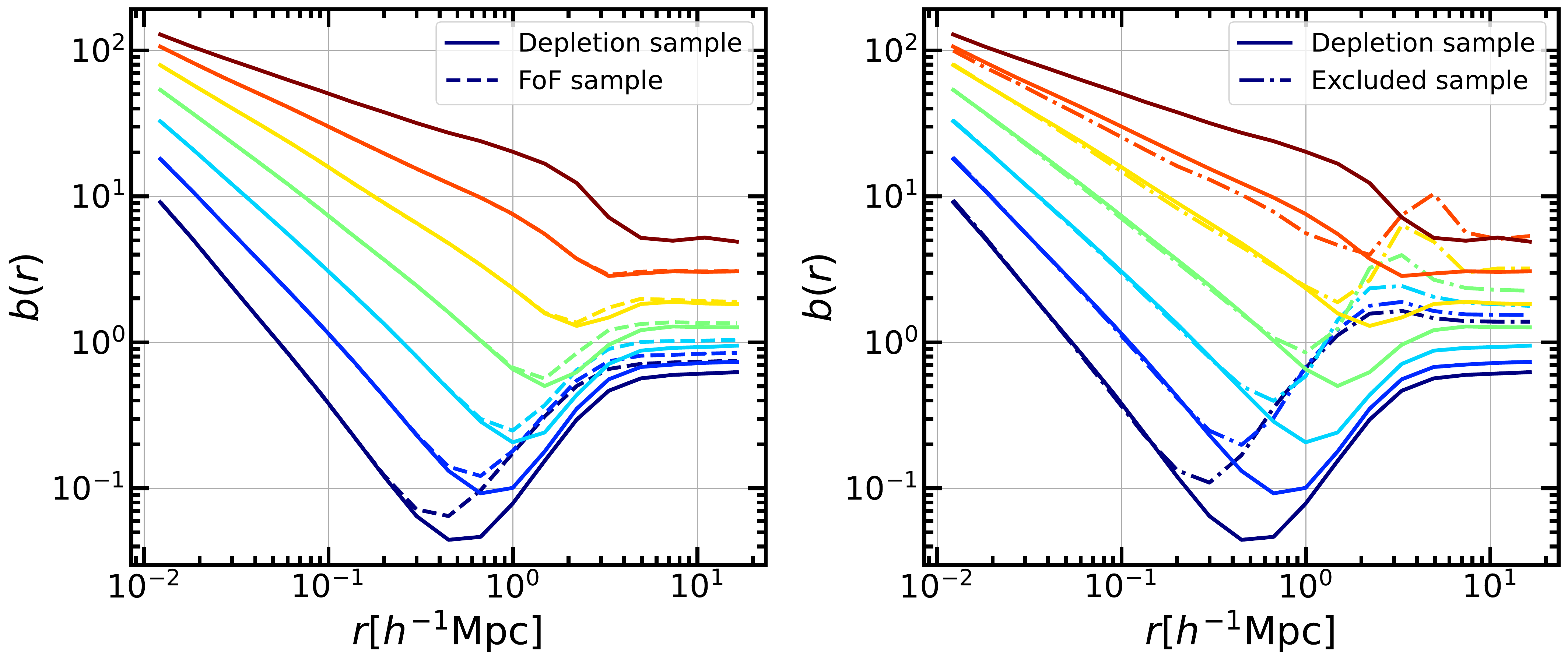}
    \caption{Bias profiles for the original FoF sample (full sample), the cleaned sample according to $R_{\rm id}$ (labelled as depletion sample), and the removed haloes due to the cleaning (labelled as excluded sample). Different colours show the profiles in different halo mass bins with the same labels as in Figure \ref{fig:r_ratio}.}
    \label{fig:samp_sepa}
\end{figure*}

%Hard-sphere exclusion scheme excludes a fraction of haloes whose boundaries touch the boundaries of more massive haloes. 
Hereafter, we refer to the original FoF catalog as the FoF sample, haloes cleaned by $R_{\rm id}$ as the depletion sample, and the difference between the two as the excluded sample. In Table~\ref{tab:halopopulation} we summarize the properties of the FoF and depletion haloes. The average bias profiles of the different samples are shown in Figure \ref{fig:samp_sepa}. In the left panel, we compare the bias profiles of the FoF and depletion sample. For massive haloes, their bias profiles are barely affected by the selection. For low mass ones, the bias profile of depletion sample has a lower bias minimum than that of the FoF sample, and the location of the minimum is also further out. At linear scales, the depletion sample shows a lower linear bias than FoF sample for the same mass bin. These differences can be understood as our exclusion scheme removes haloes with massive neighbours and results in a less clustered sample at the intermediate and large scales. \citet{garcia2019halo} has discussed that different exclusion schemes affect the statistics of the halo catalog, and their conclusions agree with our results that a stricter exclusion scheme would remove haloes that are smaller and more clustered.  %It shows that low-mass haloes are easier to be captured into the accretion region of more massive haloes. Notably, these removed haloes are treated as host haloes in $S_{\rm F}$, but their accretion histories are highly affected by high-mass neighbours. In $S_{\rm I}$, they are treated as subhaloes, reflecting the ideal of depletion boundary: an isolated halo is not only virialized but also dominates the material accretion at its local environment. 

The profiles of the excluded haloes are shown in the right panel, which all show a peak feature after the bias trough and before reaching the linear bias. This peak reflects the existence of the massive neighbour and resembles the peak found for early-forming haloes in \citep{fong2021natural}. These excluded haloes are also more clustered on large scale. We summarise the properties of excluded haloes in Appendix \ref{app:appA} and leave more detailed studies of them to future works.

% At the end of this section, we summarize the information for the seven mass bins of $S_{\rm F}$ and $S_{\rm I}$. Throughout the rest of the paper, we explore the statistical properties of different halo masses based on the data in Table \ref{tab:halopopulation}. 

\begin{table*}
    \centering
    \begin{tabular}{l|l|l|l|l|l|l|l}
         Halo Population & ${\rm log}_{10}(M/{\rm M}_{\odot})$ & $N_{\rm p}$ (Fs) & $N_{\rm p}$ (Ds) & $b$ (Fs) & $b$ (Ds) & $R_{\rm id}[h^{-1}{\rm kpc}]$ (Fs) & $R_{\rm id}[h^{-1}{\rm kpc}]$ (Ds)\\
         \hline
         MB1 & [11.50, 12.05) & 1359703 & 1138828 & 0.73 & 0.60 & 346.05($2.1R_{\rm vir}$) & 346.43($2.1R_{\rm vir}$)\\
         MB2 & [12.05, 12.60) & 445168 & 383808 & 0.82 & 0.71 & 548.91 & 493.83\\
         MB3 & [12.60, 13.15) & 140487 & 125014 & 1.02 & 0.92 & 859.56 & 831.64\\
         MB4 & [13.15, 13.70) & 40957 & 37503 & 1.37 & 1.28 & 1295.70 & 1291.01\\
         MB5 & [13.70, 14.25) & 10008 & 9466 & 1.95 & 1.90 & 1883.01 & 1923.55\\
         MB6 & [14.25, 14.80) & 1624 & 1581 & 3.10 & 3.07 & 2664.53 & 2733.43\\
         MB7 & [14.80, 15.35) & 103 & 103 & 4.97 & 4.97 & 3802.53 & 3860.34\\
    \end{tabular}
    \caption{Halo populations in the FoF sample (Fs) and the depletion sample (Ds). $N_{\rm p}$ represents the number of haloes of given mass bin. $b$ is the linear bias estimated by averaging the bias profile in radial range of $5-20h^{-1}{\rm Mpc}$. $R_{\rm id}$ is the inner depletion radius measured from the velocity profile, with the exception of $R_{\rm id}$ for MB1 which is estimated from $R_{\rm id}\simeq 2.1R_{\rm vir}$.}
    \label{tab:halopopulation}
\end{table*}

\section{A halo model including the exclusion effects}
\label{sec:sec3}

\subsection{Analytical framework}
\subsubsection{The classical approach}
Our goal in this subsection is to represent the halo- matter correlation function using the physical quantities associated with the halo. The key idea of the halo model is that the total density field can be equivalently described by the superposition of the matter distribution in individual haloes~\citep[see][for a comprehensive review]{cooray2002halo}. Let $\textbf{\textit{x}}_i$ be the position of the $i$th halo in a given halo population, the halo number density field can be written as 
\begin{align}
n(\textbf{\textit{x}})&=\sum_i \delta^{\rm D}(\textbf{\textit{x}}-\textbf{\textit{x}}_i) \nonumber \\
&= \sum_m \sum_k \delta^{\rm D}(\textbf{\textit{x}}-\textbf{\textit{x}}_k|m)
\label{eq:halo_field}
\end{align}
where the $\textbf{\textit{x}}_k$ represents the position of the $k$th halo in $m$th mass bin, and $\delta^{\rm D}(\textbf{\textit{x}})$ is the Dirac delta function.

With the internal matter distribution of each halo given by its density profile $\rho_{\rm h}(\textbf{\textit{r}})$, the total density field can be written as
\begin{align}
\rho(\textbf{\textit{x}})&= \sum_i \rho_{\rm h(}\textbf{\textit{x}}-\textbf{\textit{x}}_i)\nonumber \\
%&=& \sum_m \sum_k \rho_h(\textbf{\textit{x}}-\vec{x_k}|m)\nonumber \\
%&=& \sum_m [n*\rho_{\rm h}](\textbf{\textit{x}}|m)\nonumber \\
&= \sum_m n(\textbf{\textit{x}}|m)\circledast \rho_{\rm h}(\textbf{\textit{x}}|m),
\label{eq:matter_field}
\end{align}
 %which is simply the halo field convolved with the density profile of each halo.
where $\circledast$ represents the convolution operation. Here we have assumed haloes of the same mass share the same density profile. However, it is straightforward to generalize the equation to more general halo profiles. %Equation \ref{eq:matter_field} demonstrates the matter field can be decomposed into the halo field convolved by the halo density profile.

With these two fields, we can calculate the two-point correlation functions between the halo and matter fields. Using Equation \ref{eq:halo_field}, the halo-halo correlation function between two samples of haloes, $h_1$ and $h_2$, can be expressed as
\begin{align}
\xi_{ h_1h_2}(\textbf{\textit{r}}) &= \langle \delta_1(\textbf{\textit{x}})\delta_2(\textbf{\textit{x}}+\textbf{\textit{r}}) \rangle\nonumber \\
%&=& \frac{\langle n_1(\textbf{\textit{x}})n_2(\textbf{\textit{x}}+\textbf{\textit{r}}) \rangle}{\bar{n}_1\bar{n}_2}-1 \nonumber \\
&= \frac{\langle \sum_{i_1}\delta^{\rm D}(\textbf{\textit{x}}-\textbf{\textit{x}}_{i_1})\sum_{i_2}\delta^{\rm D}(\textbf{\textit{x}}+\textbf{\textit{r}}-\textbf{\textit{x}}_{i_2}) \rangle}{\bar{n}_1\bar{n}_2}-1 \nonumber \\
&= \frac{\bar{n}_{12}\delta^{\rm D}(\textbf{\textit{r}})}{\bar{n}_{1}\bar{n}_{2}}+\frac{\sum_{i_1\neq i_2}\delta^{\rm D}(\textbf{\textit{x}}_{i_1}-\textbf{\textit{x}}_{i_2}+\textbf{\textit{r}})}{\bar{n}_1\bar{n}_2V} -1,\label{eq:xi_hh}
\end{align}
where $\bar{n}_{12}$ represents the number density of shared haloes among the two halo samples. To derive the last equality, we have replaced the ensemble average with volume average under the assumption of ergodicity, $<\cdot>=\frac{1}{V}\int \cdot\ud^3r$.
The first term arises from the correlation of discrete haloes with themselves and is only non-zero at $\textbf{\textit{r}}=0$, known as the 1-halo term. The second and third terms can be recognized as the 2-halo term, $\xi_{\rm hh}^{\rm 2h}$, representing the average overdensity of neighbouring haloes at separation $\textbf{\textit{r}}$ around each of the $h_1$ halo.%One can interpret the two-point correlation function as the ensemble-averaged overdensity of population $h_2$ around population $h_1$, i.e. $\xi_{\rm h_1h_2}(\textbf{\textit{r}})= \langle \delta_{h_2}(\textbf{\textit{r}}|h_1) \rangle$. In this case, the first term represents the contribution of the central halo and the remaining terms form the reduced halo-halo correlation function $\xi^{2h}_{\rm hh}(\textbf{\textit{r}})$, representing the contribution from the neighbouring haloes.

Starting from Equation~\eqref{eq:matter_field}, the matter overdensity field can be rewritten as the weighted sum of the halo overdensity field convolved with halo profiles, as
\begin{equation}
    \delta_{\rm m}(\textbf{\textit{x}})= \sum_m f(m) \delta(\textbf{\textit{x}}|m)\circledast u_{\rm h}(\textbf{\textit{x}}|m),
\end{equation} where 
\begin{equation}
f(m)=\frac{\bar{n}(m)}{\bar{\rho}}\int \rho_{\rm h}(\textbf{\textit{r}}|m)\ud^3 r\label{eq:mass_frac}
\end{equation}
is the fractional contribution from haloes of mass $m$ to the mean density, and 
\begin{equation}
u_{\rm h}(\textbf{\textit{r}})=\frac{\rho_{\rm h}(\textbf{\textit{r}}|m)}{\int \rho_{\rm h}(\textbf{\textit{r}}|m)\ud^3 r} \label{eq:rho_norm}
\end{equation}
is the normalized halo density profile.\footnote{We explicitly keep the mass integral $\int \rho_{\rm h}(\textbf{\textit{r}}|m)\ud^3 r$ in Equations~\eqref{eq:mass_frac} and \eqref{eq:rho_norm}, to allow for the use of a general mass label, $m$, that is not necessarily equal to the mass integral. See Section~\ref{sec:mass_func} for more detail.} It then follows that the halo-matter correlation function between the matter field and a halo population with a given mass, $M$, is 
\begin{align}
\xi_{\rm hm}(r|M) &= \langle \delta_{\rm h}(\textbf{\textit{x}}|M)\delta_{\rm m}(\textbf{\textit{x}}+\textbf{\textit{r}}) \rangle \nonumber \\
%&=& \frac{\langle n_h(\textbf{\textit{x}}|M)\rho(\textbf{\textit{x}}+\textbf{\textit{r}}) \rangle}{\bar{n}_h(M)\bar{\rho}(m)}-1 \nonumber \\
%&=& \frac{\langle n_h(\textbf{\textit{x}}|M)\sum_m[n*\rho](\textbf{\textit{x}}+\textbf{\textit{r}}) \rangle}{\bar{n}_h(M)\bar{\rho}(m)}-1 \nonumber \\
%&=& \frac{\sum_m\bar{n}_h(M)\bar{n}(m)[1+\xi_{\rm hh}(\textbf{\textit{r}}|m,M)]*\rho(\textbf{\textit{r}})}{\bar{n}_h(M)\bar{\rho}_m}-1 \nonumber \\
%&=& \frac{\bar{\rho}_m+\sum_m\bar{n}(m)\xi_{\rm hh}(\textbf{\textit{r}}|m,M)*\rho(\textbf{\textit{r}}) }{\bar{\rho}_m}-1 \nonumber \\
%&=& \frac{1}{\bar{\rho}_m}\sum_m\bar{n}(m)\xi_{\rm hh}(\textbf{\textit{r}}|m,M)*\rho(\textbf{\textit{r}}) \nonumber \\
&= \sum_m f(m)\xi_{\rm hh}(r|m,M)\circledast u(r).
\end{align}
Replacing the summation with the integral over mass and substituting in Equation~\eqref{eq:xi_hh}, we get
\begin{align}
\xi_{\rm hm}(r|M) %&=& \frac{1}{\bar{\rho}_m}\int_0^{\infty} \bar{n}(m)\xi_{ hh}(\textbf{\textit{r}}|m,M)\circledast \rho_h(\textbf{\textit{r}}|m)\ud m \nonumber \\
 &=\xi^{\rm 1h}_{\rm hm}(r|M) + \xi^{\rm 2h}_{\rm hm}(r|M),
\label{eq:hm-decomp}\\
%\end{eqnarray}
%where the corresponding 1-halo and 2-halo terms are given by
%\begin{eqnarray}
    \xi^{\rm 1h}_{\rm hm}&= \frac{\rho_{\rm h}(r|M)}{\bar{\rho}_{\rm m}} \label{eq:hm-1halo} \\
\xi^{\rm 2h}_{\rm hm}&= \int f(m)\xi^{\rm 2h}_{\rm hh}(r|m,M)\circledast u_{\rm h}(r|m)\ud m , \label{eq:exact_2h}
%&=& \frac{1}{\bar{\rho}_m}\int_0^{\infty} \bar{n}(m)\xi^{2h}_{ hh}(\textbf{\textit{r}}|m,M)\circledast \rho_h(\textbf{\textit{r}}|m)\ud m . \nonumber \\
\end{align} where $\xi^{\rm 1h}_{\rm hm}$ and $\xi^{\rm 2h}_{\rm hm}$ are the 1-halo and 2-halo terms respectively.
In most previous works, the 2-halo term is often simplified with the linear approximation that haloes trace the matter density field at large scales with a constant bias,
\begin{equation}
    \xi^{\rm 2h}_{\rm hh}(r|m,M)\approx b(m)b(M)\xi_{L}(r)
    \label{eq:approximation}
\end{equation}
where the $b(m)$ is the linear halo bias and $\xi_L(r)$ is the linear correlation function. The inner structure of haloes becomes unimportant at these scales so that halo profiles can be approximated as point masses, $u_{\rm h}(\textbf{\textit{r}}|m)\approx \delta^{\rm D}(\textbf{\textit{r}})$. Further with the help of the local mass conservation relation
\begin{equation}
    \int f(m) b(m) \ud m=1,\label{eq:local_conser}
\end{equation} the 2-halo term becomes
\begin{equation}
\xi^{\rm 2h}_{\rm hm}(r|M) \approx b(M)\xi_L(r).\label{eq:linear_corr}
\end{equation}

This approximation provides a convenient way to calculate the 2-halo term, but it also leaves some issues. First of all, the linear approximation \ref{eq:approximation} only holds at large scales but becomes more and more problematic on smaller scales. In addition, the density profile of neighbouring haloes needs to be carefully considered at intermediate scales rather than being replaced by point masses. Moreover, the halo-halo correlation function can not extend to the interior of the central halo by construction. These problems that occur on intermediate scales have been generalized as the ``halo exclusion effect'' \citep{cooray2002halo}, which is essentially due to the unclear understanding of the halo boundary and the inappropriate extrapolation of the linear correlation function. Calculating the exact 2-halo term must return to the equation \ref{eq:exact_2h}.

\subsubsection{Modifying the 2-halo term }
Equation \ref{eq:exact_2h} illustrates that the determination of the 2-halo term involves three quantities: the density profile $\rho(r)$, the halo mass function $\bar{n}(m)$, and the halo-halo correlation function $\xi_{\rm hh}(r|m,M)$. The former two have been extensively studied, while the latter has received less attention especially around the exclusion scale. Previous studies paid more attention to the behaviour of the halo-halo correlation function at large scales, which is close to a power law \citep[e.g.,][]{correlationDavis, correlationJing98, correlationJing02, zehavi2004departures}. For a given halo catalog, the exact behavior of the halo-halo correlation around the halo boundary can be sensitive to how the halo-finder distinguishes between haloes and subhaloes in detail. In the context of the halo model, it is also not guaranteed that any halo-finding scheme may be able to yield a self-consistent halo catalog for the model. %is related to the artificial definition of the halo substructure so that no definitive solutions. 
Based on our depletion halo sample, we specifically discuss the characteristics of the halo distribution in Section \ref{sec:xihh} and give a parameterized formula to accurately describe the halo-halo correlation functions down to the depletion radius. Exploiting this formula the 2-halo term can be directly calculated without any approximation at intermediate scales.

There are some additional theoretical uncertainties in directly evaluating 
Equation \ref{eq:exact_2h}. For example, the lower mass limit of haloes cannot reach 0 due to freestreaming of CDM particles \citep[e.g.,][]{Green05,Profumo06,Schneider13} . %is no way to directly verify the lowest mass of dark matter haloes due to limitations in observational accuracy and simulation resolution; 
On the other hand, simulations can only resolve haloes above a much higher mass limit than the freestreaming scale, leaving large uncertainties in extrapolating the halo properties to lower masses. %there is a large degree of variation in the results of different theoretical models extrapolated to $m = 0$. 
Fortunately, the internal structure of the very low mass neigbouring halo is no longer important when evaluating the 2-halo term, as long as their sizes are well below the boundary scale we are modelling. As a result, we can treat haloes below a certain mass threshold, $m_{\rm res}$, as point masses, and model them with an unresolved halo term. Rewriting equation \ref{eq:exact_2h} we obtain
\begin{align}
\xi^{\rm 2h}_{\rm hm}(r|M) &= \xi^{\rm res}_{\rm hm}(r|M) +\xi^{\rm unr}_{\rm hm}(r|M), \label{eq:calcu_1}\\
%\xi_{hm}(\textbf{\textit{r}}|M) &=& \xi^{1h}_{hm}(\textbf{\textit{r}}|M) + \xi^{\rm res}_{hm}(\textbf{\textit{r}}|M) +\xi^{\rm unr}_{hm}(\textbf{\textit{r}}|M) \\
%\label{eq:calcu_1h}
%\xi^{1h}_{hm}(\textbf{\textit{r}}|M) &=& \frac{\rho_h(\textbf{\textit{r}}|M)}{\bar{\rho}_m} \\
\xi^{\rm res}_{\rm hm}(r|M) &= \int_{m_{\rm res}}^{m_{\rm max}} \xi^{\rm 2h}_{\rm hh}(r|m,M)\circledast u_{\rm h}(r|m) f(m)\ud m ,\label{eq:calcu_res} \\
\xi^{\rm unr}_{\rm hm}(r|M) &= \int^{m_{\rm res}}_{m_{\rm fs}} \xi^{\rm 2h}_{\rm hh}(r|m,M)f(m)\ud m+ f_{\rm d} \xi_{\rm hd}(r|M),
\label{eq:calcu_2}
\end{align}
where $m_{\rm max}$ is the maximum halo mass, and $m_{\rm fs}$ is the freestreaming mass. To be more general, we have added a diffuse matter term, with $f_{\rm d}$ representing the fraction of diffuse matter in the universe, and $\xi_{\rm hd}(r|M)$ is the correlation function between the halo sample and the diffuse matter. Together with haloes below $m_{\rm res}$, they form the unresolved mass term, $\xi^{\rm unr}_{\rm hm}$. Note $m_{\rm res}$ does not have to correspond to the actual halo mass resolution of the simulation, but represents the mass above which we can reliably model their mass function and density profile, and below which they can be safely approximated as point sources.  %$\xi^{\rm res}_{hm}$ and $\xi^{\rm unr}_{hm}$ thus are contributions from resolved neighbouring haloes and the unresolved mass respectively. 

With the above decomposition, the remaining task is to model $\xi^{\rm 2h}_{\rm hh}$ and $\xi_{\rm hd}$. As we have abandoned the linear approximation, we will measure $\xi^{\rm 2h}_{\rm hh}$ directly from our simulation data. As we show in Section~\ref{sec:xihh} below, we find the $\xi^{\rm 2h}_{\rm hh}$ for haloes of different masses all share a universal shape. This enables us to easily parameterize the function for arbitrary halo masses, and generalize it to the diffuse matter limit to obtain $\xi_{\rm hd}$.

%As we have abandoned the linear approximation in our new derivation, the unresolved term is still not numerically solvable. Alternatively, we assume that all haloes follow a similar spatial distribution, so that the halo-halo correlation function for haloes below the $m_1$ can be decomposed as $\xi_{hh}(r|m, M)=B(m, M)\zeta(r)$, where the universal halo distribution (UHD) $\zeta(r)$ is independent to the $m$ and $M$ but only depends on $r$. This new assumption appears to be identical to the linear approximation however, there is a fundamental difference in their scale of application and the specific forms. We assume that UHD $\zeta$ is applicable at scales outside the halo boundary and is equal to -1 inside the halo, representing that no neighbouring halo resides within the central halo. The linear approximation only holds at large scales and assumes that haloes trace the linear correlation function that extends to the interior of the halo, which causes the unphysical extrapolation at intermediate and large scales. By replacing the linear approximation with UHD assumption, the unresolved term can eventually be rewritten to a simple form. We will illustrate the reasonableness of this assumption for depletion samples using the simulation data and parametric formula in Section \ref{sec:xihh}, and give the specific numerical method in Section \ref{sec:result1}.

It is important to emphasize that since we reconstruct the halo-matter correlation function using the depletion sample, the density profile and the halo mass function need to be revised. In other words, we need to seek a set of halo model components $\{\rho(r)$, $n(m)$, $\xi_{\rm hh}(r|m_1,m_2)\}$ that are self-consistent with the definition of the depletion sample. We put more details in the next three subsections. %There is a high degree of freedom in adjusting parameters and functions in the halo model, so any arbitrary definition of the halo boundary may also have corresponding model components that can accurately reconstruct the halo-matter correlation function, even though such a boundary has no clear physical meaning. In other words, the decomposition of the density field is mathematically a multi-solution problem.

\subsection{Halo-halo correlation function}
\label{sec:xihh}

\subsubsection{Measurements}\label{sec:xihh_data}
Under spherical average, the halo-halo correlation function (Equation~\eqref{eq:xi_hh}) between two mass-selected halo populations can be rewritten in a more practical form as %for given halo populations can be measured by,
\begin{equation}
    \xi_{\rm hh}(r|m,M)=\frac{\langle n(r,m)|M\rangle}{\bar{n}(m)}-1
\end{equation}
where $m$ and $M$ represent the masses of the two populations, and $\langle n(r,m)|M\rangle$ is the average number density of mass $m$ haloes at a separation $r$ around haloes of mass $M$. Note the correlation is commutable in the masses, with $\xi_{\rm hh}(r|m,M)=\xi_{\rm hh}(r|M,m)$. When $m=M$, $\xi_{\rm hh}(r,m)$ becomes the auto-halo correlation function.
\begin{figure*}
	\includegraphics[width=\textwidth]{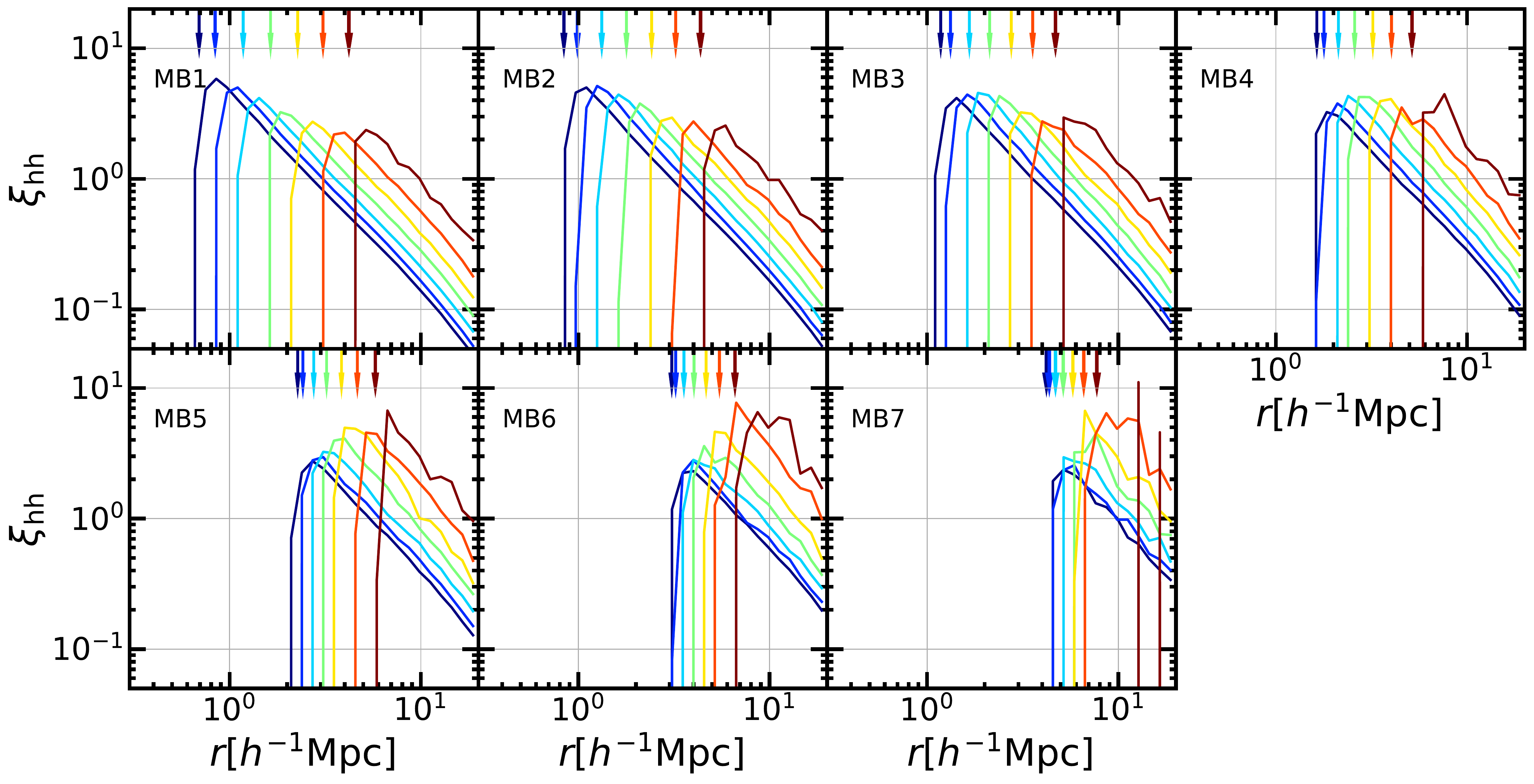}
    \caption{Halo-halo correlation functions for different central and neighbouring haloes. Each panel shows the correlation functions around a given central mass bin, and different colours represent mass bins of neighbouring haloes as in Figure \ref{fig:r_ratio}. Arrow marks the location of the exclusion scale, which is defined as $r_{\rm t}=R_{\rm id}^{\rm cen}+R_{\rm id}^{\rm nei}$.}
    \label{fig:CFhh}
\end{figure*}

Figure \ref{fig:CFhh} shows the halo-halo correlation functions with different central and neighbouring masses in our depletion sample. One of the most significant features of $\xi^{\rm 2h}_{\rm hh}(r|m, M)$ is that a sharp truncation appears at 
\begin{equation}
r_{\rm t}=R_{\rm id}(m_1)+R_{\rm id}(m_2),\label{eq:r_t}
\end{equation} reflecting the exclusion criterion we used in selecting isolated haloes. % small scales, whose position varies with $m$ and $M$. This is expected as we have applied the hard-sphere exclusion rule, to select haloes with $d_{12}>R_{\rm ex} = R_{\rm id}(m_1)+R_{\rm id}(m_2)$. By marking the location of $R_{\rm id}(m_1)+R_{\rm id}(m_2)$, we find that the exclusion radius is very close to the truncation, which agrees with our expectations. 
 %, which follows a power-law as we mentioned later. Furthermore, the correlation functions of different halo populations are similar, suggesting that haloes with different masses follow a universal distribution at large scales. 
 Beyond the exclusion scale, $\xi^{\rm 2h}_{\rm hh}(r|m, M)$ appears to follow a universal shape, which differs from either the linear or non-linear matter correlation function. On the linear scale, however, $\xi^{\rm 2h}_{\rm hh}(r|m,M)$ is expected to reduce to the linear approximation in Equation~\eqref{eq:approximation}. This means we can parametrize the non-linear halo-halo correlation function with the help of a scale-dependent function, $\zeta(r)$, through
\begin{equation}
    \xi^{\rm 2h}_{\rm hh}(r|m,M)= b(m)b(M)\zeta(r)\xi_{\rm mm}(r),
    \label{eq:CFhh}
\end{equation}
with $\zeta(r)\rightarrow 1$ on the linear scale. On smaller scales, higher order biases become significant, and $\zeta(r)$ encodes the non-linear effects in halo clustering \citep{van2013cosmological}. %The dependence on $r_{\rm t}$ depicts the truncation due to halo exclusion. 
Note that the radial bias function is independent of the halo masses except for a truncation at $r_{\rm t}$ which we will model later.%, reflecting the similarity of halo clustering. 

\begin{figure*}
	\includegraphics[width=\textwidth]{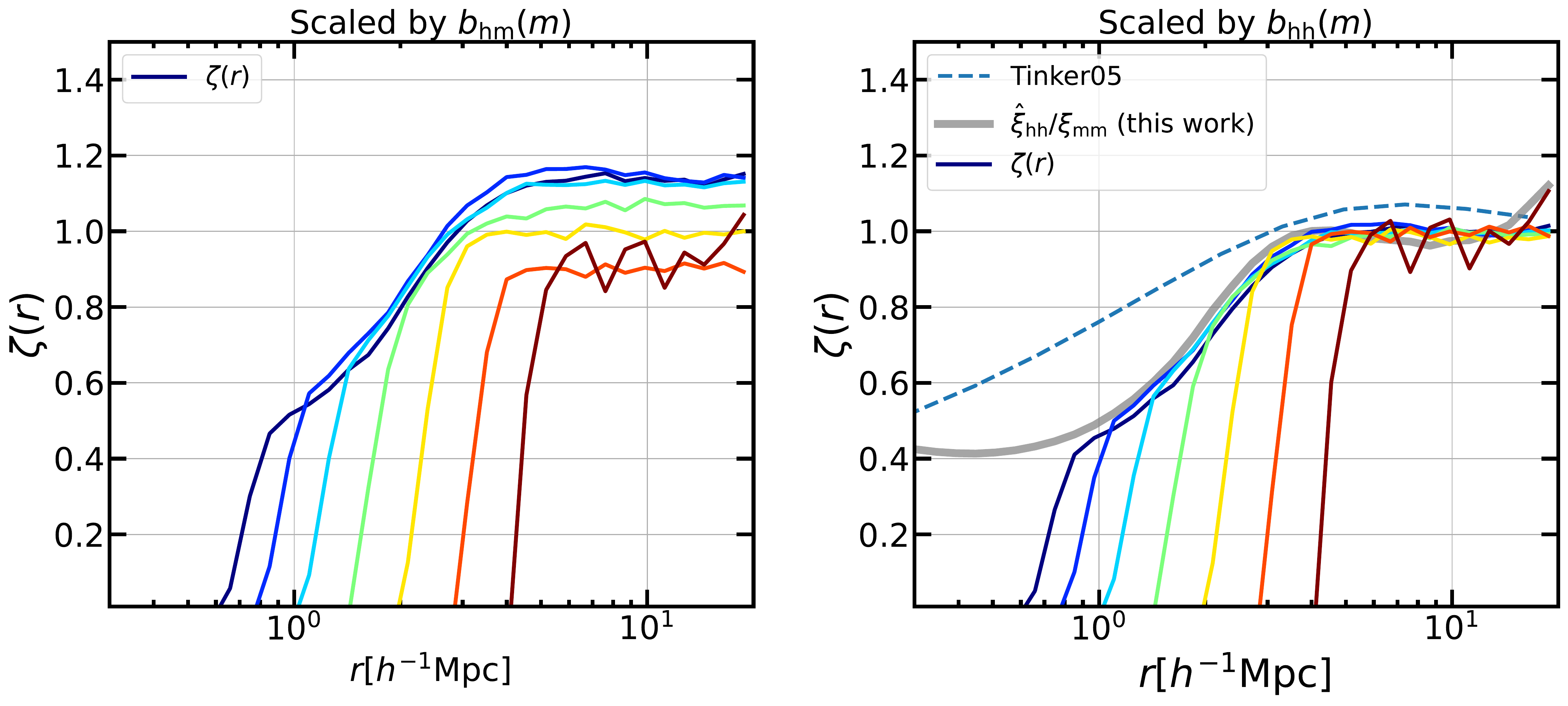}
    \caption{The radial bias functions measured from different halo populations. The central halo mass is $10^{11.5}h^{-1}{\rm M}_{\odot}<M<10^{12.05}h^{-1}{\rm M}_{\odot}$ (MB1). The colours of solid lines represent the neighbouring halo mass with the same labels as in Figure \ref{fig:r_ratio}. The dashed blue line is the result from Equation~\ref{eq:zeta}. The thick gray line is the unit halo correlation scaled by $\xi_{\rm mm}(r)$ from Equation~\ref{eq:xi_hh_star}. %\jx{remove $\hat{\xi}$ from left panel}. 
    The y-axis of the left and right panel represent the radial bias functions scaled by $b_{\rm hm}(m)$ and $b_{\rm hh}(m)$, respectively.}
    \label{fig:zeta_comp}
\end{figure*}

%Figure~\ref{fig:zeta_comp} compares the measured $\zeta(r)$ with the above fitting function. 
To extract $\zeta(r)$ in our halo sample, we tried two ways to estimate the linear bias, through 
\begin{equation}
b_{\rm hm}(m)=\xi_{\rm hm}(r|m)/\xi_{\rm mm}(r)
\end{equation}
and 
\begin{equation}
    b^2_{\rm hh}(m)=\xi_{\rm hh}(r|m,m)/\xi_{\rm mm}(r)
\end{equation} respectively. On the linear scale the above estimates are expected to become scale-independent.
%The radial bias function can be obtained by,
%\begin{equation}
%    \zeta(r)=\frac{\xi_{hh}(r|m_1,m_2)}{b_{hm}(m_1)b_{hm}(m_2)\xi_{mm}}
%    \label{eq:zeta_estimator}
%\end{equation}
%where $b$ is the linear halo bias, which can be estimated independently from halo-matter and halo-auto correlation functions: $b_{\rm hm}(r|m)=\xi_{\rm hm}(r|m)/\xi_{\rm mm}(r)$ and $b^2_{\rm hh}(r|m)=\xi_{\rm hh}(r|m)/\xi_{\rm mm}(r)$. 
More specifically, we calculate the bias profiles $b_{\rm hm}(r|m)$ and $b_{\rm hh}(r|m)$ and average them in radii of $5-20 h^{-1}{\rm Mpc}$ to obtain the corresponding bias estimators $b_{\rm hm}$ and $b_{\rm hh}$. The radial bias function is then obtained from Equation~\eqref{eq:CFhh} with the estimated linear biases and the measured correlation functions. Using different bias estimators to calculate the radial bias function, we find significant deviations in results shown in Figure \ref{fig:zeta_comp}. For $b_{\rm hm}$, the resulting radial bias functions diverge from each other on the linear scale, suggesting that $b_{\rm hm}$ does not correctly reflect the clustering in the halo-halo correlation function. When replacing $b_{\rm hm}$ with $b_{\rm hh}$, all results shown in the right panel converge to 1 on the linear scale. Furthermore, the measured $\zeta(r)$ profiles indeed become mass independent across scales, except for their different truncation radii. %a clear similarity between the radial bias functions can be observed at linear and trans-linear scales. All results follow an envelope curve, and for different halo populations, $\zeta(r)$ truncates at different exclusion radii. For comparison, we also plot $\zeta_T(r)$ in each panel.

\citet{tinker2005mass} proposed to model $\zeta(r)$ with the following fitting function,
\begin{equation}
    \zeta_{\rm T}(r,z)=\frac{[1+1.17\xi_{\rm mm}(r,z)]^{1.49}}{[1+0.69\xi_{\rm mm}(r,z)]^{2.09}},
    \label{eq:zeta}
\end{equation} which is also plotted in both panels for comparison. This fitting function fails to match our measurements. On the linear scale, it does not converge to 1. At smaller scales, $\zeta_{\rm T}(r)$ also deviates from our measurements in the depletion sample and extends below the exclusion scale, which is unphysical for a real halo distribution. 

The fact that the $\zeta(r)$s converge to various values in the left panel of Figure \ref{fig:zeta_comp} reflects that $b_{\rm hm}$ is not identical to $b_{\rm hh}$. In fact, $b_{\rm hh}$ is sensitive to the first and second-order bias, while $b_{\rm hm}$ only responds to the first-order bias (see Appendix \ref{app:hobias}). It means that $b_{\rm hh}$ is a more suitable choice when modelling the halo-halo correlation function. Based on the above findings, we will establish a parametric formula to describe the halo-halo correlation function in the next step using $b_{\rm hh}(m)$ and the similarity of the halo distribution.

\subsubsection{Fitting Formula}
\label{sec:CFhh_fit}
Equation~\eqref{eq:CFhh} can be rewritten as
\begin{equation}
    \xi_{\rm hh}^{\rm 2h}(r|m,M)= b(m)b(M)\hat{\xi}_{\rm hh}(r),
    \label{eq:CFhh2}
\end{equation} where we have introduced a unit halo-halo correlation, $\hat{\xi}_{\rm hh}(r)$, to describe the common shape of the correlation. This unit correlation can be interpreted as the auto-correlation function for a linearly unbiased halo population, with $b(m_\ast)=1$.

The unit halo correlation $\hat{\xi}_{\rm hh}(r)$ is related to the radial bias function through $\hat{\xi}_{\rm hh}(r)=\zeta(r)\xi_{\rm mm}(r)$. It is thus equivalent to fit $\xi^{\rm 2h}_{\rm hh}$ through either function. However, $\hat{\xi}_{\rm hh}(r)$ is much simpler compared to $\zeta(r)$, as can be seen from Figures~\ref{fig:CFhh} and \ref{fig:zeta_comp}. More fundamentally, in the halo model framework, haloes are the building blocks of the universe, while the total matter field is a derived quantity. It is thus more consistent to start from $\hat{\xi}_{\rm hh}(r)$ as a model input, rather than relying on $\zeta(r)$ which is defined relative to the matter field. 

We find the $z=0$ unit halo correlation can be well described by a power-law,
\begin{equation}
    \hat{\xi}_{\rm hh}(r)=\left(\frac{r}{r_0}\right)^{-\gamma},\label{eq:xi_hh_star}
\end{equation} with best-fitting parameters $r_0=4.96$ and $\gamma=1.58$ in our halo sample. In Figure \ref{fig:zeta_comp} we plot the unit halo correlation scaled by $\xi_{\rm mm}(r)$. Compared with $\zeta_{\rm T}(r)$, our parametric formula successfully captures the shapes of the radial bias functions at non-linear scales for different neighbouring masses. On linear scale, our model slightly deviates from 1 and shows a growing trend beyond $\sim 10h^{-1}{\rm Mpc}$. It implies that extending the power-law formula to very large scale is potentially problematic. However, the problem can be easily solved by joining our model with the classical linear model (see Equation~\eqref{eq:linear_corr}) which works well on large scale.% \yf{This problem is beyond our main scope. We recommend readers interested in larger scales to use other models, such as the classical linear theory}.%However, this problem can be easily solved by joining our unit correlation function with the shape of the linear correlation function on these very large scales. %lvable. On the one hand, the matter distribution at larger scales can be well described by linear theory; on the other hand, the halo-matter correlation at intermediate scales can be accurately reconstructed using the unit halo correlation shown in later sections, which proves that the power-law description is valid at the scales of our interest.}

By further modelling halo exclusion with a step function, the complete halo-halo correlation function can be modelled as
\begin{equation}
    1+\xi_{\rm hh}^{\rm 2h}(r|m_1,m_2)=[1+b_{\rm hh}(m_1)b_{\rm hh}(m_2) \hat{\xi}_{\rm hh}(r)]\mathrm{H}(r-r_{\rm t}),\label{eq:xi_hh_fit}
\end{equation} where $\mathrm{H}(x)$ is the Heaviside step function which is unity at $x>0$ and 0 otherwise, and $r_{\rm t}$ is given by Equation~\eqref{eq:r_t}. Note the step function is multiplied to $1+\xi$ instead of $\xi$, to ensure zero density within $r_{\rm t}$.%\jx{or still use 1+xi=(1+xi)*H?}

The above fitting formula can be easily generalized to describe the diffuse matter distribution, 
\begin{equation}
    1+\xi_{\rm hd}(r|M)=[1+b_{\rm hh}(M)b_{\rm d} \hat{\xi}_{\rm hh}(r)]\mathrm{H}(r-r_{\rm t}),
\end{equation} 
where $b_{\rm d}$ is the bias of the diffuse matter. For both unresolved haloes and the diffuse matter, $r_{\rm t}\simeq R_{\rm id}(M)$ as the size of the neighbouring halo or diffuse particle is negligible. The unresolved correlation term, Equation~\eqref{eq:calcu_2}, then becomes
\begin{align}
    \xi^{\rm unr}_{\rm hm}(r|M) = b_{\rm hh}(M)b_{\rm unr}\hat{\xi}_{\rm hh}(r)\mathrm{H}(r-r_{\rm t}) \nonumber \\
    +f_{\rm unr}[{\rm H}(r-r_{\rm t})-1],
    \label{eq:fit_unr}
\end{align}
where 
\begin{equation}
    b_{\rm unr}=\int_{m_{\rm fs}}^{m_{\rm res}} b_{\rm hh}(m) f(m)\ud m+ f_{\rm d} b_{\rm d}
    \label{eq:A_def}
\end{equation}
and 
\begin{equation}
    f_{\rm unr} = \int_{m_{\rm fs}}^{m_{\rm res}}f(m)\ud m+f_{\rm d},
\end{equation}
%\jx{I've combined $f_u$ and $f_d$ into $f_{unr}$ to be consistent with the definition of unresolved term. Please check if other occurrences are consistent. Also for submission to MN please use British English: e.g., neighbour instead of neighbour.}
are the effective bias and mass fraction of the unresolved term respectively. Mathematically, the latter term of Equation~\eqref{eq:fit_unr} equals $-f_{\rm unr}$ at small scales, grows rapidly near $r_{\rm t}$, and converges to 0 eventually at large scales. This term arises because of the disappearance of neighbours in the inner region of the central halo. It enforces the correlation function equal to $-1$ at $r\lesssim R_{\rm id}$ and has no contribution to the correlation function at large scales. However, the inner region of the halo is too dense so it does not matter whether the halo-halo correlation equal to 0 or $-1$. In other words, we can neglect the contribution of this term in the calculation of the unresolved term. Finally we obtain
\begin{equation}
    \xi^{\rm unr}_{\rm hm}(r|M) = b_{\rm hh}(M)b_{\rm unr}\hat{\xi}_{\rm hh}(r)\mathrm{H}(r-R_{\rm id}(M)),
    \label{eq:fit_unr2}
\end{equation}

Even though we cannot measure $b_{\rm d}$ and $b_{\rm hh}$ for the unresolved haloes directly, it is possible to infer $b_{\rm unr}$ from the local mass conservation relation, Equation~\eqref{eq:local_conser}. %One can easily generalize Equation~\eqref{eq:local_conser} in presence of diffuse matter, to obtain
Rewritting Equation~\eqref{eq:A_def},
\begin{align}
    b_{\rm unr}&= \int b_{\rm hh}(m) f(m)\ud m-\int_{m_{\rm res}}^{m_{\rm max}} b_{\rm hh}(m) f(m)\ud m,\label{eq:A_theo_exact}\\
    &\approx 1-\int_{m_{\rm res}}^{m_{\rm max}} b_{\rm hh}(m) f(m)\ud m,\label{eq:A_theo}
\end{align} which means $b_{\rm unr}$ can be fully determined using the properties of the resolved haloes, and is not a free parameter. However, Equation~\eqref{eq:A_theo} is only an approximate relation, because $b_{\rm hh}$ is not the same as the linear bias required in Equation~\eqref{eq:local_conser} as we discussed in Section~\ref{sec:xihh_data}. As we discuss in Appendix~\ref{app:mass_conser}, we can still expect that the parameter $b_{\rm unr}$ is constant if mass conservation is held. To model the halo-matter correlation with high accuracy and maintain the mass conservation of our model, we will leave $b_{\rm unr}$ as the only free parameter for a global fit in all mas bins. We will discuss $b_{\rm unr}$ in more detail in Section~\ref{sec:result1} and Appendix~\ref{app:mass_conser}. 
% The fitting formula can be further derived to describe the unresolved term. Considering the distribution of unresolved haloes follows UHD, the halo-halo correlation can be rewritten as
% \begin{eqnarray}
%     \xi_{\rm hh}(r|m,M) &=& \{B(m,M) \left( \frac{r}{r_0}\right)^{-\gamma}\}\Phi(r, r_t) \nonumber \\
%     &+& \{\Phi(r, r_t)-1\}
% \end{eqnarray}
% where $M$ and $m$ are masses of central haloes and unresolved neighbouring haloes. Assuming unresolved haloes are regarded as point masses, the truncation scale can be approximated as $r_t\simeq R_{\rm id}(M)$. Mathematically, the latter term $\Phi(r, r_t)-1$ equals $-1$ at small scales, grows rapidly near $r_t$, and converges to 0 eventually at large scales.This term arises because of the disappearance of neighbours in the inner region of the central halo. It enforces the correlation function equal to $-1$ at $r\lesssim R_{\rm id}$ and has no contribution to the correlation function at large scales. However, the inner region of the halo is too dense so it does not matter whether the halo-halo correlation equal to 0 or $-1$. In other words, we can neglect the contribution of term $\Phi(r, r_t)-1$ in the calculation of the unresolved term. By integrating the function with respect to $r$, we finally obtain,
% \begin{equation}
%     \xi^{\rm unr}_{\rm hm}(r|M) = A(M)\left( \frac{r}{r_0}\right)^{-\gamma}\Phi(r, R_{\rm id}(M))
%     \label{eq:fit_unr}
% \end{equation}
Setting the parameters $r_0$ and $\gamma$ as best-fitting values, 
$b_{\rm unr}$ is the only parameter to be optimized.

In summary, combining Equations~\eqref{eq:calcu_res}, \eqref{eq:xi_hh_star}, \eqref{eq:xi_hh_fit} and \eqref{eq:fit_unr2} we can calculate the resolved and unresolved terms, with $b_{\rm unr}$ as a free parameter to fit for.

\subsection{Halo bias}\label{sec:bias}
\begin{figure}
    \includegraphics[width=\columnwidth]{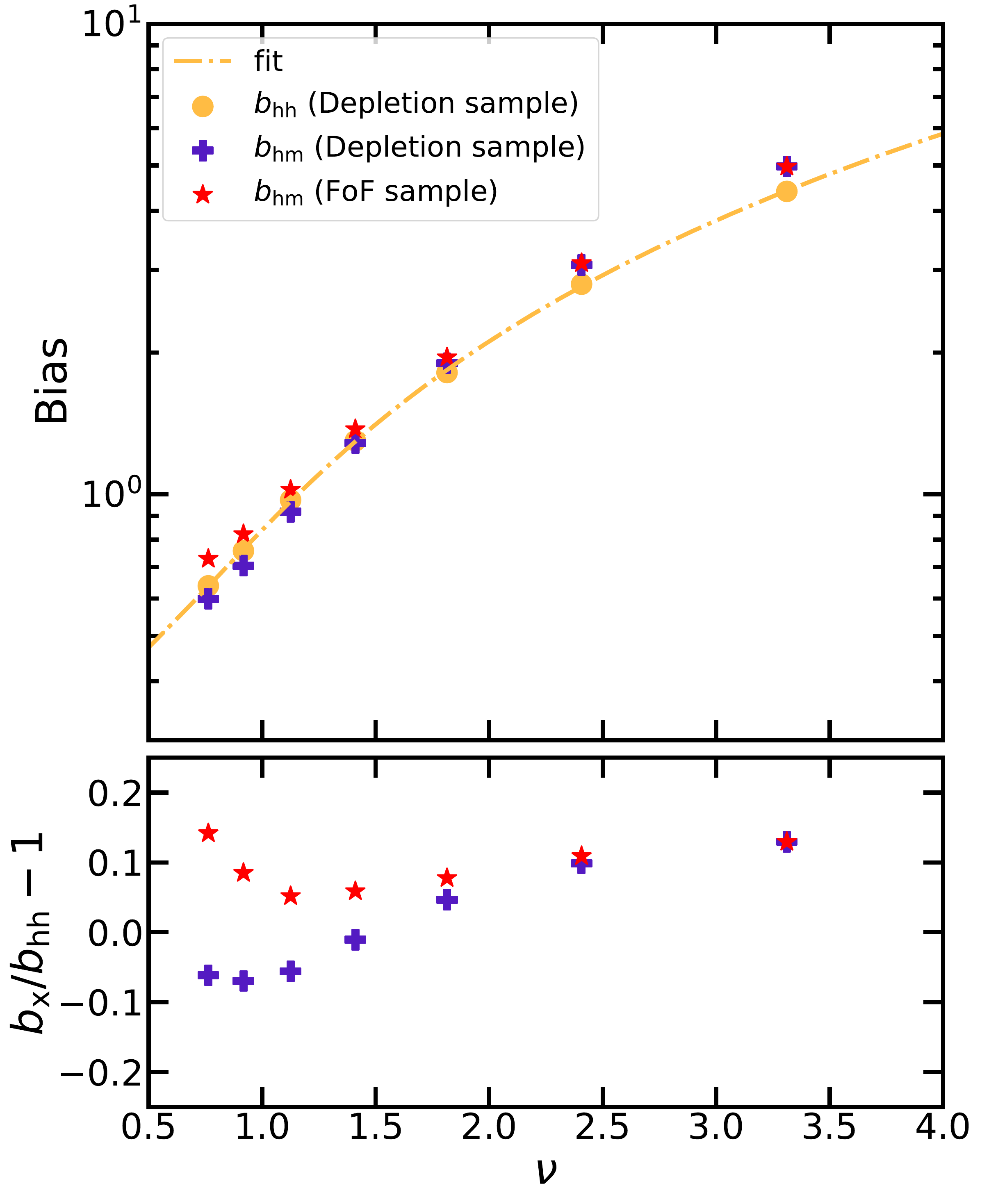}
    \caption{The halo bias. \textit{Top}: Dots and crosses represent the bias estimators $b_{\rm hh}$ and $b_{\rm hm}$, respectively, on the depletion sample. Dashed line is the fit to $b_{\rm hh}$ using Equation~\eqref{eq:bias}. Red stars represent the $b_{\rm hm}$ from the FoF sample. \textit{Bottom}: relative deviations between $b_{\rm hm}$ (depletion and FoF sample) and $b_{\rm hh}$ (depletion sample).}
    \label{fig:halobias}
\end{figure}

Since we measure halo biases from the depletion sample, no existing models can accurately reproduce our results over the mass range we considered. We adopt the functional form suggested by \citet{jing1998accurate} %in fitting the ${b}_{\rm hh}(m)$,  
to fit the measured bias in our sample with
\begin{equation}
    b_{\rm hh}(m) = \left[\frac{0.5}{\sigma^4(m)}+1\right]^c[1+D\sigma^d(m)+E],
    \label{eq:bias}
\end{equation}
where $\sigma(m)$ is the variance of the linear density field within a top-hat filter containing mass $M$. % defined as $\sigma(M)=\delta_c/\nu (M)$. 
We implement a fit using $b_{\rm hh}$s from the depletion sample,  and the best-fitting parameters for our results are,
\begin{align}
    c &= 0.206, \nonumber \\ 
    d &= 1.494, \nonumber \\ 
    D &= 0.731, \nonumber \\
    E &= -0.959 .
\end{align}
We use this new fitting formula to predict the halo bias for neighbouring haloes when calculating the 2-halo term. The bias data and the fit of $b_{\rm hh}$ are shown in Figure \ref{fig:halobias}. For comparison, we also plot the $b_{\rm hm}$ from the depletion and FoF samples respectively. The relative deviation between $b_{\rm hm}$ and $b_{\rm hh}$ for the depletion sample is negative at the low mass end and positive at the high mass end, with a crossing mass found in MB4 ($M\sim 10^{13.4}h^{-1}{\rm M}_{\odot}$). The $b_{\rm hm}$ measured from the two samples differ in low mass bins but become identical at the high mass end, consistent with the analysis for the bias profiles in Figure \ref{fig:samp_sepa}. 

\subsection{Halo density profile}
\label{sec:dens_prof}
We use the Einasto formula \citep{einasto1965construction,einasto1969andromeda} to fit the halo density profile from simulation,
\begin{equation}
    \rho_{\rm EIN}(r) = \rho_{\rm s} {\rm exp}\left( -\frac{2}{\alpha}\left[ \left(\frac{r}{r_{\rm s}}\right)^\alpha-1 \right]\right)
    \label{eq:EIN}
\end{equation}
In most previous works \citep[e.g.,][]{EinastoNavarro,gao2008redshift}, Einasto profiles were applied in the radial range $r<R_{\rm vir}$. Some studies \citep[e.g.,][]{prada2006far,EinPrada2012} have tried to extend the Einasto profile to fit the outskirts of haloes up to around $2R_{\rm vir}$ which is very close to the inner depletion radius, and they found that it works for low-mass haloes as well as some relaxed high-mass haloes. 
In this work, we use the Einasto profile to fit halo profiles using the data within $R_{\rm id}$.  
%In most previous studies~\citep[e.g.,][]{einasto1965construction,einasto1969andromeda}, this formula has been adopted to fit the density profile in the radial range $r<R_{\rm vir}$. However, Some studies also have found that the Einasto profile can be extended to fit the outer profile up to $2-3R_{\rm vir}$ \citep{prada2006far}. This upper limit corresponds approximately to the depletion radius of the halo. In other words, the halo profile inside the depletion region can be well described by the Einasto profile alone. Such a universal profile is also consistent with the results of \citet{fong2021natural} who found that the size and density at the depletion radius show simple scaling relations with respect to the virial quantities. In this work, we fit the density profile with the Einasto formula in the radial range $r<R_{\rm id}$. 

\begin{figure}
    \includegraphics[width=\columnwidth]{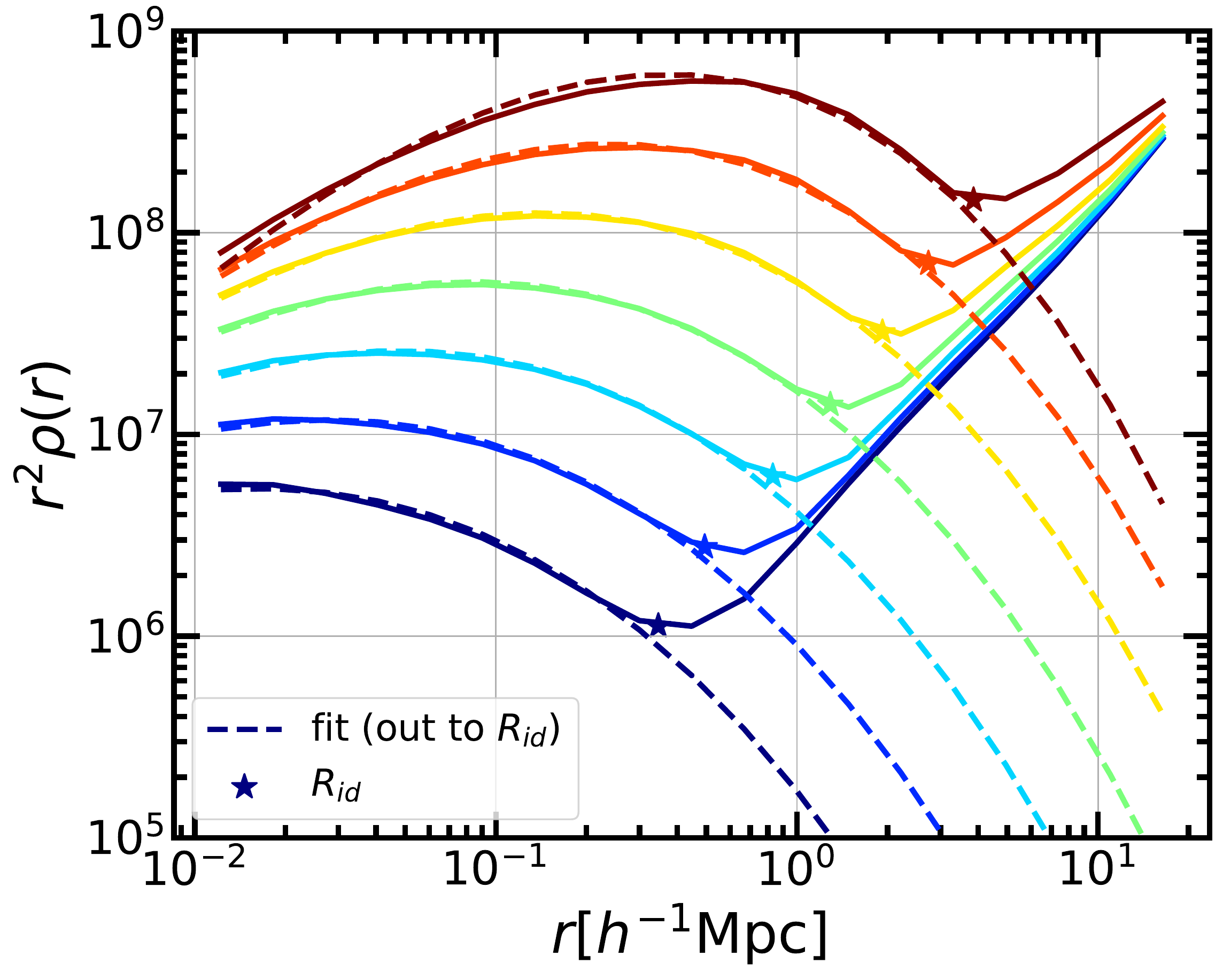}
    \caption{Reduced fits to density profiles inside the inner depletion radii. Solid lines are total density profiles of different mass bins. Dashed lines are fitting results of the Einasto formula without truncation. We only input the data of radial bins smaller than $R_{\rm id}$ and fix parameter $\alpha$ according to Equation \ref{eq:alpha} to optimize other parameters of Einasto formula.}
    \label{fig:EIN_prof}
\end{figure}

Note that some of the parameters are correlated in describing the halo profile. For example, \citet{gao2008redshift} developed a simple formula to describe the  $\nu-\alpha$ relation,
\begin{equation}
    \alpha = 0.155+0.0095\nu^2,
    \label{eq:alpha}
\end{equation}
where $\nu(M)=\delta_{\rm sc}/\sigma(M)$ is the peak height. %\jx{not $(\delta/sigma)^2?$, also see MassFunc below}. %It implies that we can reduce the number of free parameters. We fix $\alpha$ when fitting the density profile inside the $R_{\rm id}$ to test the reliability of the $\nu-\alpha$ relation for extended fits. 
Fixing $\alpha$ according to this relation, we find the halo profile can still be well fitted by the Einasto profile out to $R_{\rm id}$ as shown in Figure~\ref{fig:EIN_prof}. The outer profiles deviate from the predictions due to the 2-halo term involved.

\begin{figure}
    \includegraphics[width=\columnwidth]{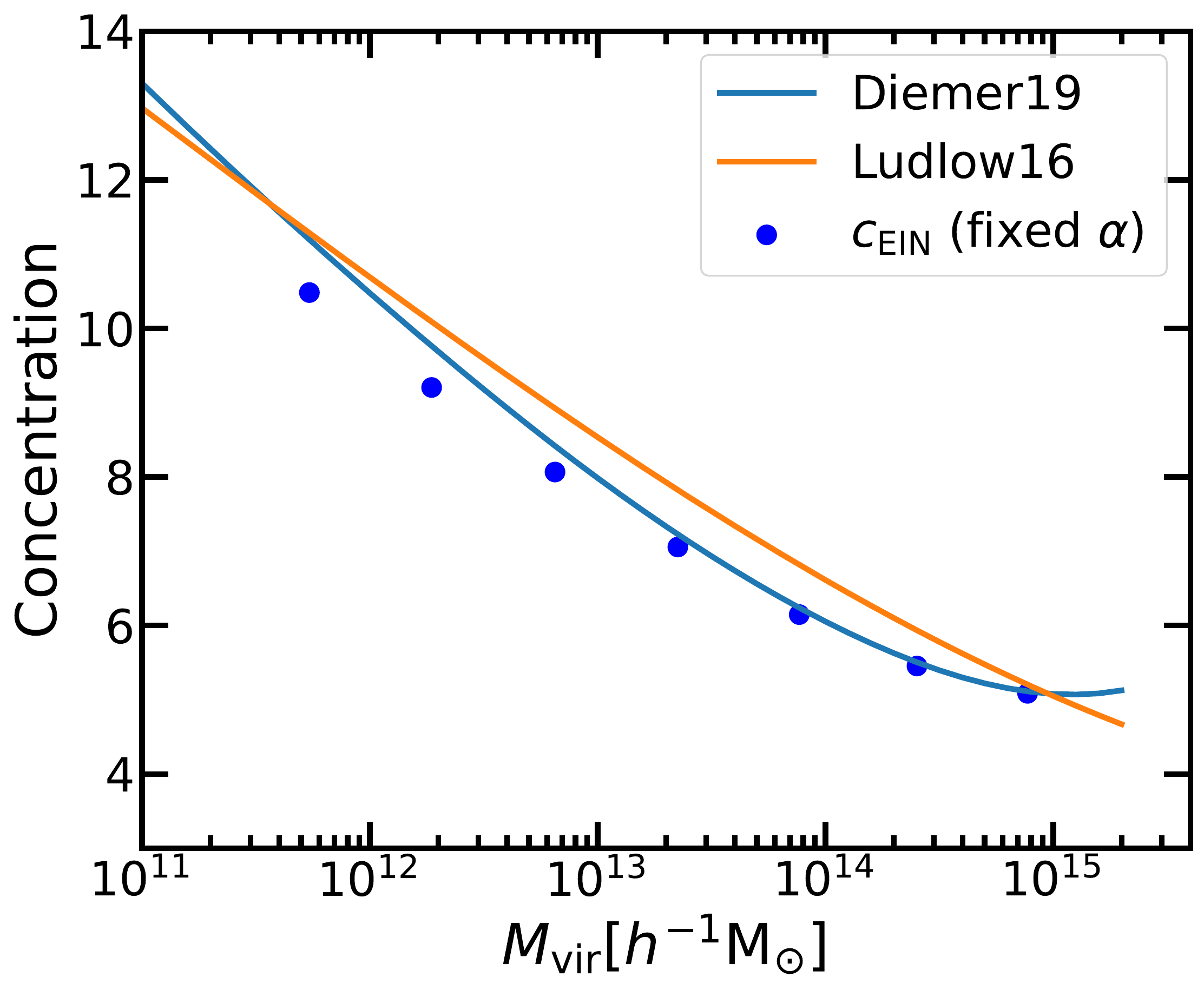}
    \caption{Mass-concentration relations from diffetent models. The solid lines are the predictions of \citet{diemer2019accurate} and \citet{ludlow2016mass}, calculated with COLOSSUS \citep{diemer2018colossus}. The dots are the concentrations of different mass bins calculated by fitting the density profile inside the $R_{\rm id}$ with fixed $\alpha$. }
    \label{fig:c_comp}
\end{figure}

With the $r_{\rm s}$ and $\rho_{\rm s}$ parameters from the reduced fits ($\alpha$-fixed), we calculate the resulting concentrations and compare them with the results from \citet{diemer2019accurate} and \citet{ludlow2016mass} in Figure~\ref{fig:c_comp}. We find that the mass-concentration relation for our depletion sample are consistent with the \citet{diemer2019accurate} result based on a FoF halo catalog especially at the high mass end. At the low mass end our result shows slightly lower concentrations, reflecting that the removed haloes tend to be more concentrated due to their proximity to a massive neighbour. Note the \citet{ludlow2016mass} result is systematically higher due to the removal of un-relaxed haloes in their measurement, compared with the result of \citet{diemer2019accurate} and ours. %This indicates that the exclusion selection hardly affects the inner profile of haloes.\jx{the removed haloes should have higher concentration as they are more tidally truncated due to the neighbour. So I expect the depletion sample to have a lower concentration, which is indeed the case compared with Ludlow. Is the Diemer19 model also subject to similar isolation selection?}

Compared with the NFW profile~\citep{navarro1995simulations,navarro1996structure,navarro1997universal}, the Einasto profile declines more rapidly in the outskirts. Despite this, it still extends well beyond the depletion radius. In other words, the depletion radius does not appear as a clear truncation in the density profile. Such an extended profile beyond the depletion radius is actually a desired feature in our model. This is because we remove FoF haloes that overlap with neighbouring more massive ones when constructing our depletion halo sample. These removed haloes do not contribute to the density field as individual haloes, but have to be counted as substructures of their massive neighbours. Besides, substructures in the original FoF halo may also extend beyond the depletion radius~\citep{2009ApJ...692..931L,2016MNRAS.457.1208H}. In principle, using the hard-sphere exclusion we may sort out the extended substructures in the halo outskirts and thus find a more accurate description for the outer profile, which we leave to future works. 

In our current model, we assume the halo profile follows the Einasto formula all the way out, which will be used in modelling both the 1-halo and 2-halo terms. Note this profile does not explicitly contain the feature of the inner depletion radius. The validity of this assumption beyond the depletion radius is supported by the good performance of the model as we show below. The success of the model also suggests that the Einasto profile happen to be a good match to the $R_{\rm id}$-based halo definition. %This assumption might be reasonable\footnote{Subhaloes arising from the accretion trace the inner density profile of the host halo, and their behavior are indistinguishable from infall particles \citep{2016MNRAS.457.1208H}. Therefore it is a natural expectation that the material to be follow the same behaviour as accretion subhaloes.} but remains to be tested. 

\subsection{Halo mass function}\label{sec:mass_func}
\begin{figure}
    \includegraphics[width=\columnwidth]{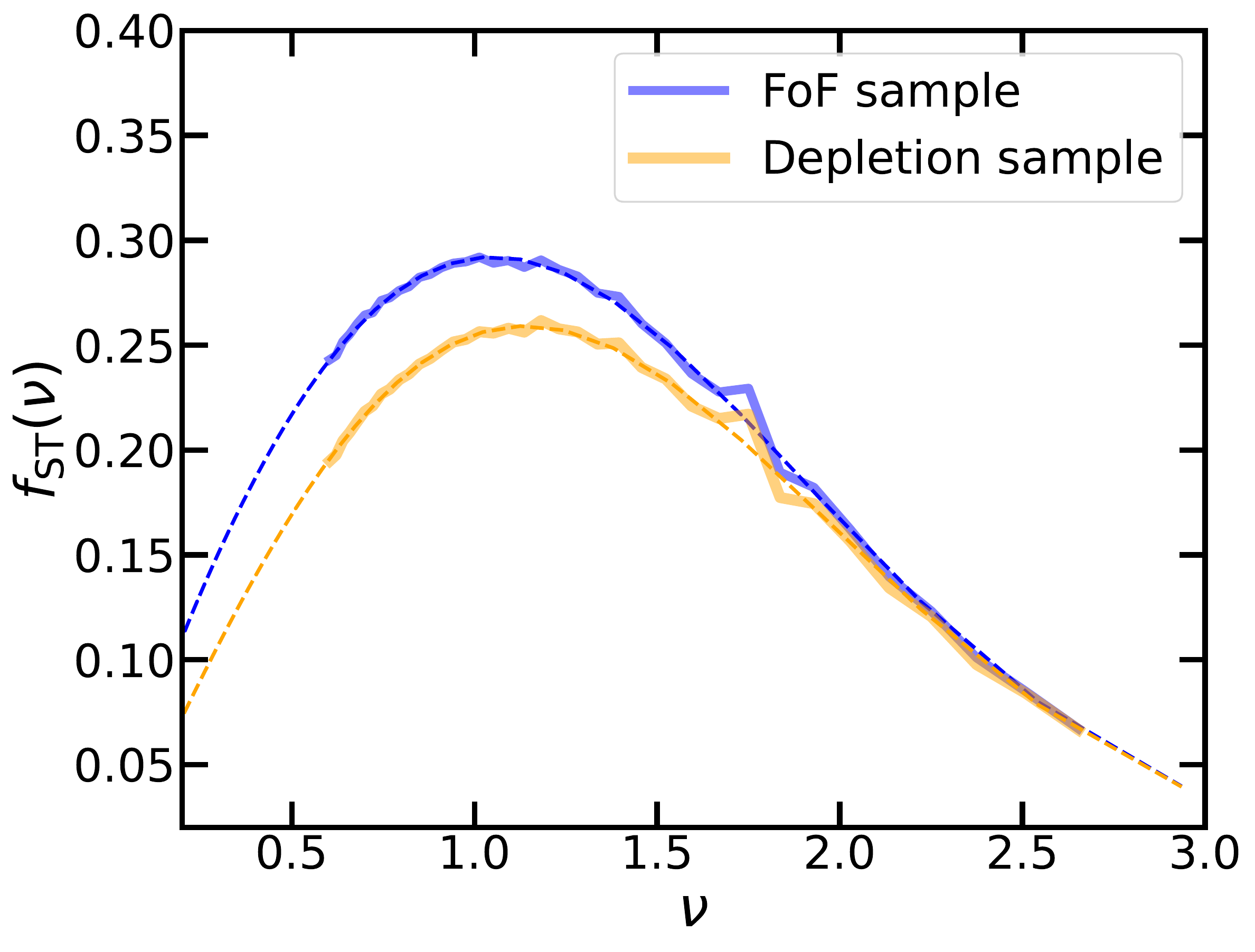}
    \caption{The halo mass function in terms of the peak height, $\nu$. The solid blue and orange lines are the mass functions for the FoF and depletion catalogs, respectively. Dashed lines show the fits with the ST formula.}%\jx{consider reducing the x-axis range, to start from 0.5 for example and use a linear scale.}\jx{use $f_{\rm ST}$ to distinguish from mass fraction $f(m)$}}
    \label{fig:HMF_iso}
\end{figure}

Figure \ref{fig:HMF_iso} compares the mass functions of the FoF sample and the depletion sample. The differences at the high mass end are negligible, while at the low mass end the depletion sample contains significantly fewer haloes due to the $R_{\rm id}$-exclusion. This result agrees with the conclusions of \citet{garcia2019halo}. We find both mass functions can be well fit with the Sheth $\&$ Tormen formula (\citealt{sheth1999large, sheth2001ellipsoidal,ST22}, hereafter ST formula),
\begin{equation}
    f_{\rm ST}(\nu)=F\sqrt{\frac{2a}{\pi}}\nu [1+(a\nu^2)^{-p}]e^{-\frac{a\nu^2}{2}},
    \label{eq:MF}
\end{equation}
where $F=0.2677$, $a=0.7765$, $p=-0.0115$ for the depletion sample. The definition of peak height, $\nu$, is the same as that in Equation~\ref{eq:alpha}. % \jx{this differs from the original format in $\nu\rightarrow \nu^2$ and $\nu f(\nu)\rightarrow f(\nu)$.} %\citet{sheth2001ellipsoidal} derives an almost identical formula by considering the ellipsoidal collapse. 
The above function is related to the halo number density through
\begin{equation}
n(M)\ud M=\frac{\bar{\rho}}{M}f_{\rm ST}(\nu)\frac{\ud \nu}{\nu}.
    % \frac{\ud n}{\ud {\rm ln}M}=f(\nu)\frac{\rho_0}{M}\frac{\ud {\rm ln}(\sigma )^{-1}}{\ud {\rm ln}M},
    \label{eq:num_dens}
\end{equation}
%where $\rho_0$ is the mean mass density of the universe, $\sigma(M)$ is the variance of the linear density field within a top-hat filter containing mass $M$, defined as $\sigma(M)=\delta_c/\nu (M)$.

According to \citet{fong2021natural}, the ratio between the mass enclosed in the inner depletion radius and the virial mass is approximately a constant, which implies that the shape of the mass function might be identical regardless of the overdensity criterion in the mass definition. To facilitate comparison with previous works, we still use the virial mass definition throughout the analysis of the mass function and mass dependence in the depletion sample. Meanwhile, we note the existence of new possible definitions of halo mass, for example, the mass enclosed in $R_{\rm id}$, or the integrated mass of the Einasto profile. We will discuss some of these definitions in the following.

\section{Results}
\label{sec:sec4}
\subsection{Model summary}
To sum up, our model describes the halo-matter correlation function through Equations~\eqref{eq:hm-decomp}, \eqref{eq:hm-1halo}, and \eqref{eq:calcu_1}, where the resolved and unresolved components are given in Equations~\eqref{eq:calcu_res}, \eqref{eq:xi_hh_star}, \eqref{eq:xi_hh_fit} and \eqref{eq:fit_unr2}, with the halo bias, density profile and mass function specified in sections~\ref{sec:bias}, \ref{sec:dens_prof} and \ref{sec:mass_func}. 

\subsection{Model performance}
\label{sec:result1}
With the model described above, we can predict the halo-matter correlation function for each halo mass bin. With $b_{\rm unr}$ estimated from Equation~\eqref{eq:A_theo}, the model components are completely fixed a priori with no free parameters. However, as we expect Equation~\eqref{eq:A_theo} is an approximate relation due to nonlinear bias, in the following we will leave $b_{\rm unr}$ as a global free parameter for all mass bins, to achieve a higher precision while maintaining mass conservation (see Appendix \ref{app:mass_conser}).

To compute the resolved term of the halo-matter correlation, we set a lower mass limit for the resolved component at $m_{\rm res}=10^{8.5}h^{-1}{\rm M}_{\odot}$, which is much lower than the minimum halo mass of $10^{11.5}h^{-1}{\rm M}_\odot$ in our halo sample. This ensures the unresolved components can be safely treated as point masses around our halo sample. The upper mass limit is set to $m_{\rm max}=10^{15.5}h^{-1}{\rm M}_\odot$, slightly larger than the maximum halo mass resolved in our simulation.

Note the Einasto profile is used in computing both the 1-halo term and the resolved term (Equation~\eqref{eq:calcu_res}). This profile involves three parameters $\rho_s$, $r_s$ and $\alpha$, which are fixed according to the $\nu-\alpha$ relation and the mass-concentration relation in Section~\ref{sec:dens_prof}. For the mass-concentration relation, it makes little difference to our model whether adopting the \citet{diemer2019accurate} relation or the results from our data in Figure~\ref{fig:c_comp}. %For a given halo mass, we compute $\alpha$ according to Equation \ref{eq:alpha} and fit the profile inside the $R_{\rm id}$ to get the parameters $\rho_s$ and $r_s$. %The halo concentration, $r_{\rm vir}/r_s$, is generated from the model of \citet{diemer2019accurate} using \textsc{COLOSSUS}. 
With $b_{\rm unr}$ as the only free parameter, we fit the model to the simulation measurements in all mass bins jointly by minimising the following merit function 
\begin{equation}
    \chi^2= \sum_j\sum_i [ \ln\xi_{\rm hm}(r_i|M_j)-\ln\xi^{\rm model}_{\rm hm}(r_i|M_j; b_{\rm unr})]^2
    \label{eq:merit_function}
\end{equation}
where $i$ represents the $i$th radial bin of the measurements and $M_j$ represents the halo mass of the $j$th mass bin. The merit function is minimized using a Python package \texttt{iminuit}\footnote{\url{https://doi.org/10.5281/zenodo.3949207}} \citep{James1975dr}. %Equation~\ref{eq:merit_function} indicates that the parameters $b_{\rm unr}$ is optimized in all mass bin with equal weights. 
%According to Appendix \ref{app:mass_conser}, we expect $b_{\rm unr}$ to be independent of halo mass. %Therefore, such a global fit automatically maintains the mass conservation in the results.  % The allowed ranges for the parameters $r_s$, $\rho_s$, $\alpha$, and $A$ in the fitting process are listed in Table \ref{tab:parameters}.
Bias profiles are obtained by scaling the resulting halo-matter correlation functions with the matter-matter correlation function $\xi_{\rm mm}(r)$.

% \begin{table}
%     \centering
%     \begin{tabular}{l|l|l}
%          Parameters &  Description & Allowed range\\
%          \hline
%          $\rho_s$ & Normalization & $[0,+\infty]$\\
%          $r_s$ & Scale radius & $[0,+\infty]$\\
%          $\alpha$ & Shape parameter & $[0.01,0.4]$\\
%          $A$ & Amplitude & $[0,1]$\\
%     \end{tabular}
%     \caption{The fitting parameters of bias profile and corresponding allowed ranges}
%     \label{tab:parameters}
% \end{table}

%We implement two schemes to fit the bias profiles. In the first scheme, we fix the three Einasto parameters and only optimize the amplitude parameter $A$. As shown in Section \ref{sec:dens_prof}, when fitting profiles in radial bins $r<R_{\rm id}$ with fixed $\alpha$, the resulting mass-concentration relation is consistent with existing models. The parameters $r_s$ and $\rho_s$ obtained by these reduced fits ensure the self-consistency of the model since the same parameter relations are used to generate the concentration and $\alpha$ when calculating the resolved term. In the single-parameter scheme, we input the Einasto parameters getting from reduced fits in Section \ref{sec:dens_prof} as fixed values in fits of the total bias profile. 

\begin{figure*}
    \includegraphics[width=\textwidth]{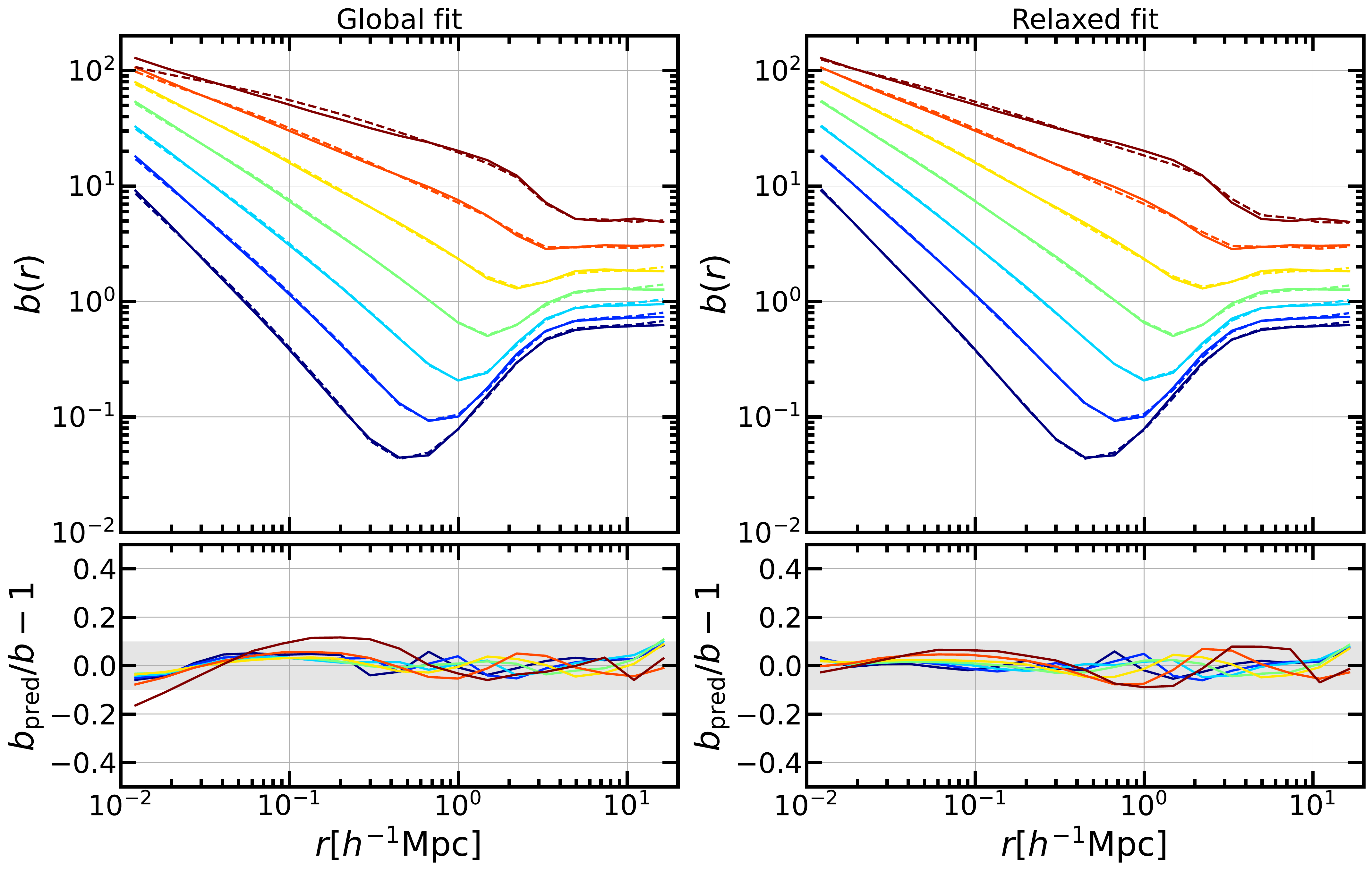}
    \caption{\textit{Left}: The bias profiles around haloes in different mass bins as labelled before. The solid lines are the measured profiles, while the dashed lines are fits with a single parameter $b_{\rm unr}$. The bottom panel shows the ratio between the data and the fits. The parameter $b_{\rm unr}$ is optimized in all mass bins jointly. \textit{Right}: Same as the left, but varying the three Einasto parameters in the 1-halo term as well $b_{\rm unr}$ for each mass bin individually during the fitting. }
    \label{fig:Results1}
\end{figure*}

The left panels of Figure~\ref{fig:Results1} show the results of fitting the above model with one free parameter, $b_{\rm unr}$, for all mass bin. Our model performs well over radial range of $0.01h^{-1}{\rm Mpc}<r<20h^{-1}{\rm Mpc}$, achieving the accuracy $\lesssim 10\%$ in all mass bins except MB7. For the highest mass bin, some systematic deviations are observed in the 1-halo component, where the model tends to under-predict the density at the smallest radius but overestimate it at $\sim 0.1 {\rm Mpc}/h$, even though it maintains good performance at intermediate and large scales. This is mostly due to the fixed $\alpha$ parameter which is not working well for the highest mass bin, as can already be seen in Figure~\ref{fig:EIN_prof}. %reflecting the parameter relations we used may exist potential issues when extrapolated to the supermassive halo. 

To see if these differences can be alleviated by a more flexible parameter combination, we further release the parameter $b_{\rm unr}$ and the Einasto parameters, $\rho_s$, $r_s$ and $\alpha$ for each mass bin during the fitting. Note Einasto profile used in the convolution of the 2-halo term is still fixed as before. This leads to a total of 4 free parameters for each mass bin. As shown in the right panels of Figure~\ref{fig:Results1}, freeing the 1-halo parameters indeed improves the fits to achieve an accuracy within 9\%  in all mass bins, implying that the Einasto profile has the ability to describe the inner profile of cluster haloes. However, these improvements come at the price of some inconsistencies between the Einasto parameters used in the 1-halo and 2-halo components, as well as inconsistencies in the mass conservation across different mass bins when $b_{\rm unr}$ varies. %Mass conservation is also not ensured in fits since $b_{\rm unr}$ is not a constant for all mass bins. %, although the corresponding parameter relations need to be carefully considered.
For simplicity, throughout the rest of the paper we choose the global fit (one free parameter) as the fiducial model, which achieves a high accuracy on intermediate and large scales we are interested in.

\begin{figure*}
    \includegraphics[width=\textwidth]{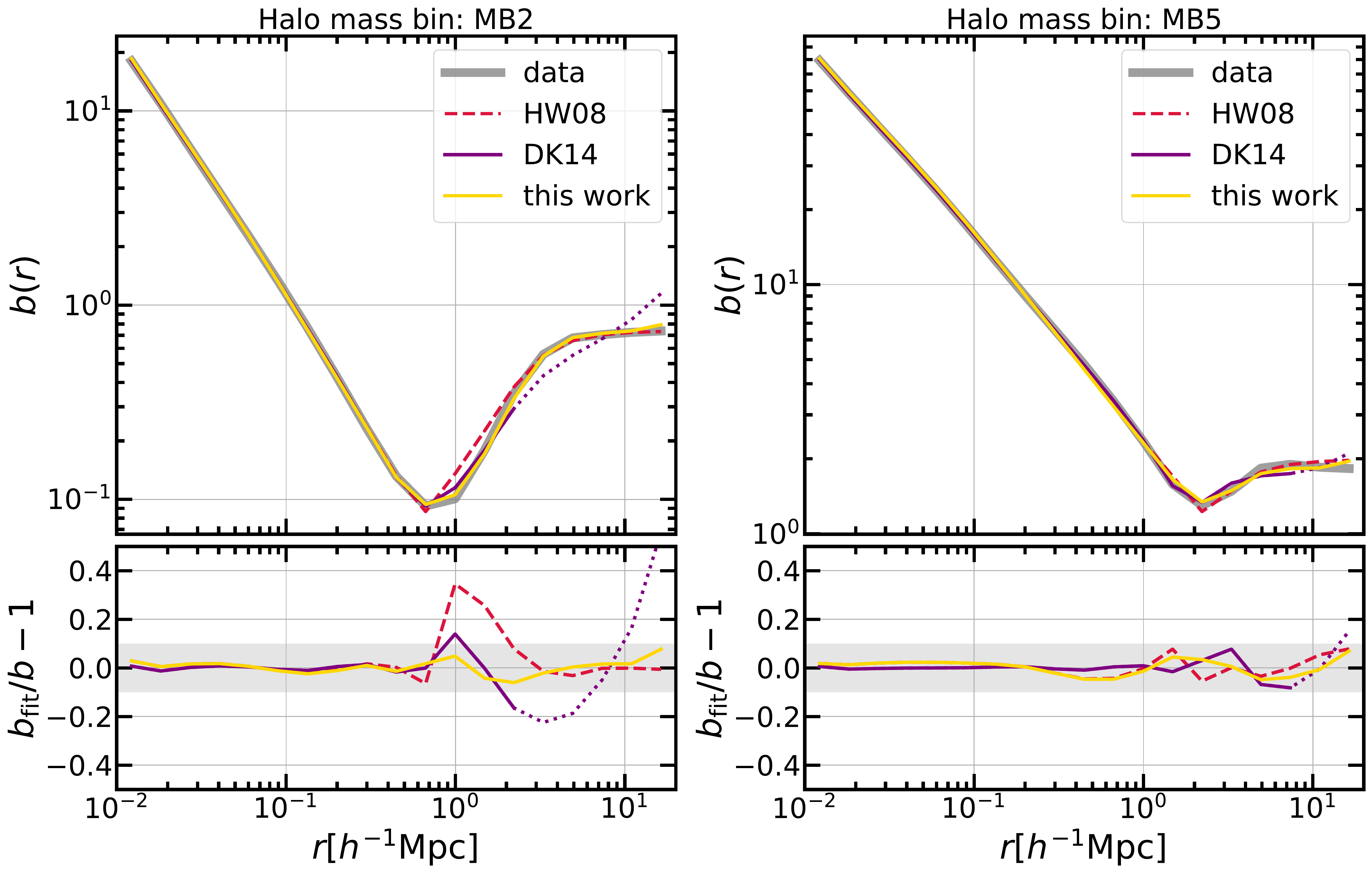}
    \caption{\textit{Left}: The bias profile fitted with different models. The grey thick line %\jx{make it thicker. it's hard to see.}
    is the total bias profile around haloes in the halo mass bin MB2 ($M\sim 10^{12.3}h^{-1}{\rm M}_\odot$). The red dashed, purple solid and yellow solid lines are the fits with the \citet{hayashi2008understanding}, \citet{diemer2014dependence}, and our model, respectively. The purple dotted line represents the DK14 model evaluated beyond its original fitting range. For a fair comparison, we relaxed the parameters of the 1-halo term in our model (as shown in the right panel in Figure \ref{fig:Results1}). All models are applied to the depletion sample. The bottom panel shows the ratio between the models and the data. The shaded region marks the 10\% deviation range. \textit{Right}: Same as the left, but fitting to a high-mass bin MB5 ($M\sim 10^{14}h^{-1}{\rm M}_\odot$).}%\jx{these are not 10\% region, but 0.1 dex!}}
    \label{fig:result_comp}
\end{figure*}

In Figure \ref{fig:result_comp} we compare our model performance with some previous works. \citet{hayashi2008understanding} (hereafter, HW08) proposed a simple model by splicing the density profile where the values of 1-halo and 2-halo terms are equal. \citet{diemer2014dependence} (hereafter, DK14) enforced a transition term on the inner profile and described the outer profile with a power law. For low-mass haloes, as shown in the left panel of figure, HW08 fits well on both small and large scales but has discontinuous and poor predictions at the intermediate scale. DK14 performs well in its range of applications ($r<9R_{\rm vir}$) but produces significant deviations on larger scales (the green dotted curve). The right panel of Figure \ref{fig:result_comp} shows the comparison in a typical high-mass bin (MB5). All of the models perform about as well as each other. Compared with these models, our model maintains a high accuracy over the entire radial range, owning to our careful and self-consistent treatment of the halo exclusion effect. Note that all models are applied to the depletion sample, although HW08 and DK14 were not originally designed based on the cleaned halo catalog. We also make a qualitative comparison with \citet{garcia2021redefinition} which developed a flexible halo model that incorporates halo exclusion by optimizing the exclusion radius with several parameters for halo masses in the range $10^{13}h^{-1}{\rm M}_{\odot}<M<10^{15}h^{-1}{\rm M}_{\odot}$. Compared to their model, our model has a wider mass range and requires much fewer fit parameters. %The differences between various models arise from the different treatments of the 2-halo term and exclusion effect.

\subsection{Contributions from different components}
\begin{figure}
    \includegraphics[width=\columnwidth]{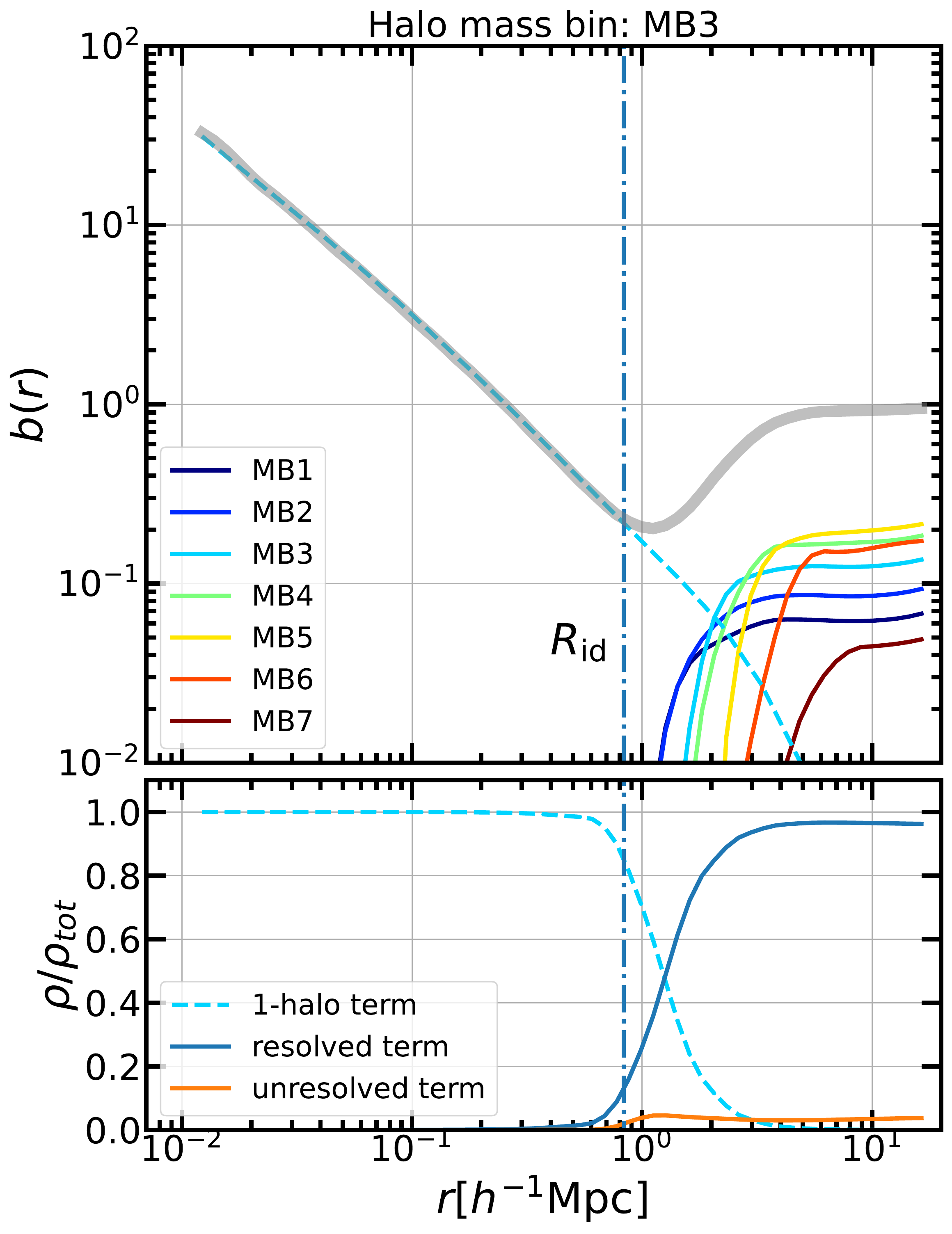}
    \caption{\textit{Top}: The decomposition of the 2-halo term around central haloes in MB3. Solid coloured lines represent the contributions of neighbouring haloes with different masses as labelled. The grey solid line and cyan dashed line represent the total bias profile and contribution from the 1-halo term, respectively. The dash-dotted blue line marks the location of $R_{\rm id}$. \textit{Bottom}: The density fractions of different components. Solid lines represent the fractions of resolved and unresolved terms. The dashed line is the fraction of the 1-halo term.}
    \label{fig:denfrac}
\end{figure}

The convolution approach allows us to study the mass contributions of neighbouring halo populations to the outer profile. To investigate the contribution from different masses of haloes, we bin neighbouring haloes by their virial masses into seven mass bins and calculate their corresponding contributions. Figure \ref{fig:denfrac} shows the decomposition of the bias profile. 

\begin{figure}
    \includegraphics[width=\columnwidth]{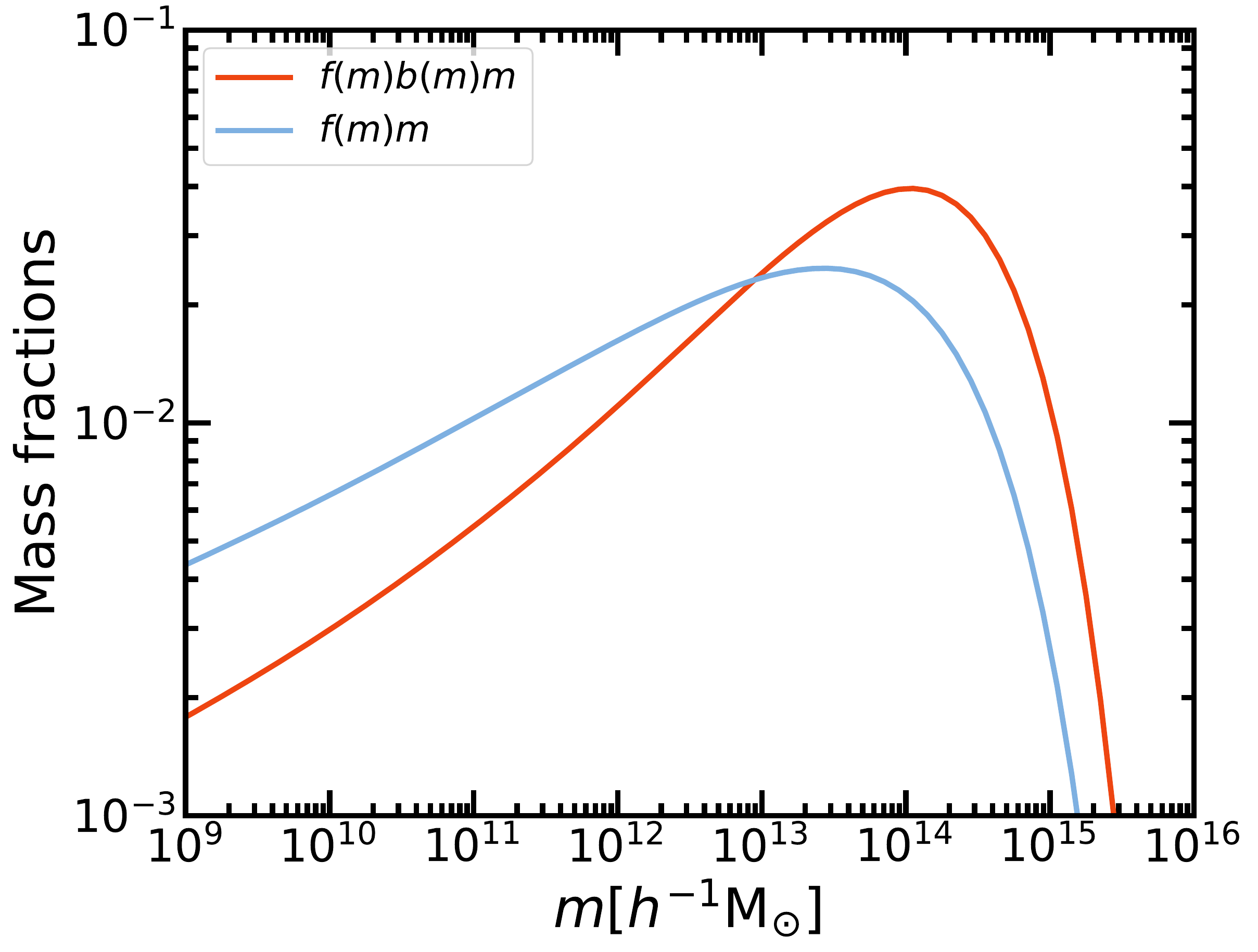}
    \caption{Mass fractions as the function of virial mass $m$. The orange curve is the biased mass fraction, while the blue curve is the total mass contribution.}
    \label{fig:fractions}
\end{figure}

On the linear scale, the contribution to the bias profile from a given logarithmic mass bin approaches a constant given by $f(m)b(m)m\ud \ln m$, which is shown in Figure \ref{fig:fractions}. %\jx{consider plotting this, and compare with $f(m)\ud m$} %At large scales where the bias profile tends to flatten out, there is a fixed ordering of the contributions from different mass bins. 
This bias fraction increases with increasing virial mass until a characteristic mass bin MB5 ($M\sim 10^{14}\msunh$), above which it decreases with halo mass. % The haloes within the characteristic mass bin contribute the most to the outer profile. For haloes with $M>10^{14.25}h^{-1}{\rm M}_{\odot}$, the bias fraction negatively correlates with the virial mass. 
This result suggests that most of the large scale clustering around haloes is contributed by group to cluster mass haloes (MB4, MB5, and MB6). Things get complicated at $1\sim 5h^{-1}{\rm Mpc}$. At these scales, the bias fraction of a given mass bin truncates near its corresponding exclusion radius, and the cutoff is smoothed due to the density profile convolved. The contribution of the low-mass haloes can extend to smaller scales because of a smaller exclusion radius so that the ordering of bias fraction is broken at these transition scales. For example, MB4 contribute the most to the bias profile at $\sim 3h^{-1}{\rm Mpc}$, and as $r$ reduce to $\sim 2h^{-1}{\rm Mpc}$, the dominant mass bin becomes MB3. In summary, low-mass haloes (MB1, MB2, and MB3) predominantly shape the outer profile of the bias trough, group to cluster mass haloes  (MB4, MB5, and MB6) contribute most of the clustered mass at the linear scale, and the most massive haloes ($M\gtrsim 10^{15}\msunh$) contribute a minor fraction on both the intermediate and larger scales due to their large exclusion radii and scarcity.


In addition to neighbouring haloes, the 1-halo term also contributes to the outer profile. We discussed the extension of the halo profile in the previous section. We use the Einasto profile without truncation to describe the 1-halo term, which implies that there is still a mass contribution from the central halo outside the halo boundary. The bottom panel of Figure \ref{fig:denfrac} shows density fractions (to avoid negative values) of the 1-halo, resolved, and unresolved terms. %\jx{how do you convert resolved and unresolve to density? together they are short of 1, but not individually}. 
The density fraction of the 1-halo term keeps unity inside $R_{\rm id}$, reflecting that the mass contribution in this region almost comes entirely from the 1-halo term. Beyond the halo boundary, the density fraction of the 1-halo term becomes comparable with that of the 2-halo term at $1-2R_{\rm id}$, which means that masses from the central and neighbouring haloes are mixed in this region. In contrast, the density fraction of the 2-halo (resolved + unresolved) term decays rapidly close to the $R_{\rm id}$ and disappears completely inside the halo. This is consistent with our definition of the halo boundary which prevents other haloes from extending into the its interior. %In the previous discussion, we mentioned that the 1-halo term needs to be extended beyond $R_{\rm id}$ to match the hard-sphere percolation. Combined with the decomposition of the bias profile, we can interpret the halo boundary in our model from another perspective: the inner depletion radius is a hard boundary for matter belonging to other haloes, preventing the 2-halo term from extending into the interior. 
The extension of the 1-halo term beyond the boundary can be regarded as the part of the halo that is still infalling, and as the matter is depleted by halo accretion from outside the $R_{\rm id}$ to build up the growing region, the 2-halo term, therefore, disappears at the boundary. This interpretation naturally reflects the physical meaning of the depletion radius suggested by \citet{fong2021natural}.

It should be emphasized that such results are, to a large extent, due to the removal of overlapping neighbouring haloes, so that using different criteria to build the halo sample will lead to different results. In principle, the decomposition of the matter density field into haloes may have multiple solutions, but not all solutions have a clear physical meaning. One advantage of our model is that it is built upon a physically defined halo boundary, i.e., the depletion radius, which separates a growing halo from the environment. %Our interpretation of the halo boundary only holds in the framework of this work. But it is still relevant for a complete understanding of the formation and evolution of dark matter haloes because the halo selection criteria of our model are based on knowledge of the depletion radius, thus the derived results self-consistently reflect the physical properties of the depletion radius.

\subsection{Characteristic radii and enclosed mass}
\begin{figure}
    \includegraphics[width=\columnwidth]{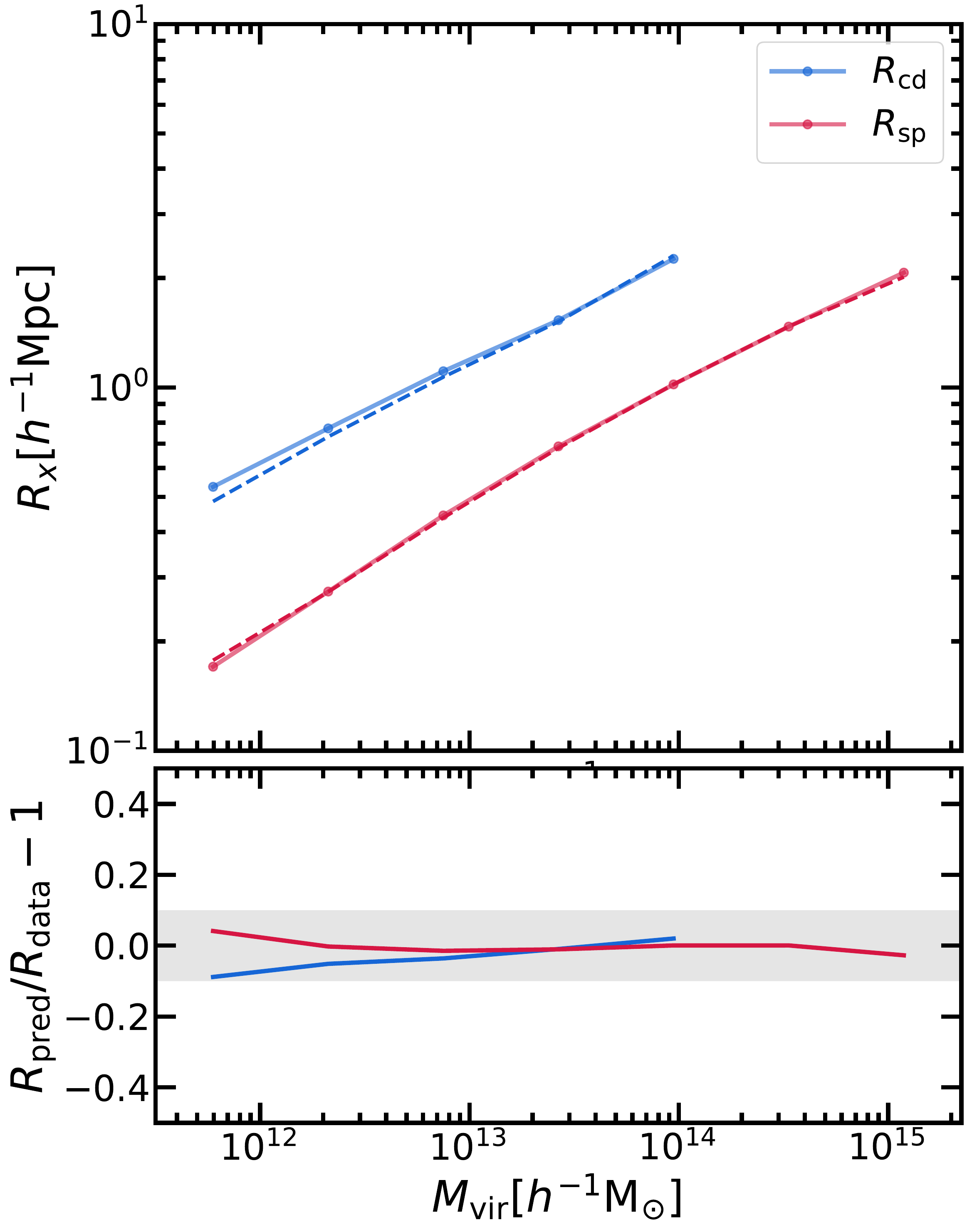}
    \caption{\textit{Top}: The $R_{\rm sp}$ and $R_{\rm cd}$ extracted from simulation profiles (solid lines and dots) and our model profiles (dashed lines). $R_{\rm cd}$s are only estimated in MB1 to MB5. \textit{Bottom}: The radius ratios of model and data. The shaded region marks the 10\% region.}
    \label{fig:Rrelation}
\end{figure}

Based on the density profile predicted by our model, we can dig out more dynamical information by measuring the characteristic radii in the profile. Here we mainly focus on the splashback radius $R_{\rm sp}$ and the characteristic depletion radius $R_{\rm cd}$. The splashback radius is measured from the steepest slope position of the density profile according to \citet{diemer2014dependence} and \citet{more2015splashback}. The characteristic depletion radius is located at the minimum of the bias trough~\citep{fong2021natural}. If we only use the Einasto formula to fit the density profile, the fitting result has no minimum in the slope profile or the bias profile. In other words, the measurement of the characteristic radii relies on both the 1-halo term and a corresponding 2-halo term. In Figure \ref{fig:Rrelation} we show the mass-radius relation of $R_{\rm sp}$ and $R_{\rm cd}$ measured from the our model profile and the simulation data. Since the characteristic depletion radius is not well defined for high-mass haloes, we only estimate $R_{\rm cd}$ in MB1 to MB5. Our model profile successfully captures these two characteristic locations with an accuracy of $\lesssim10\%$ over the mass range considered. 

\begin{figure}
    \includegraphics[width=\columnwidth]{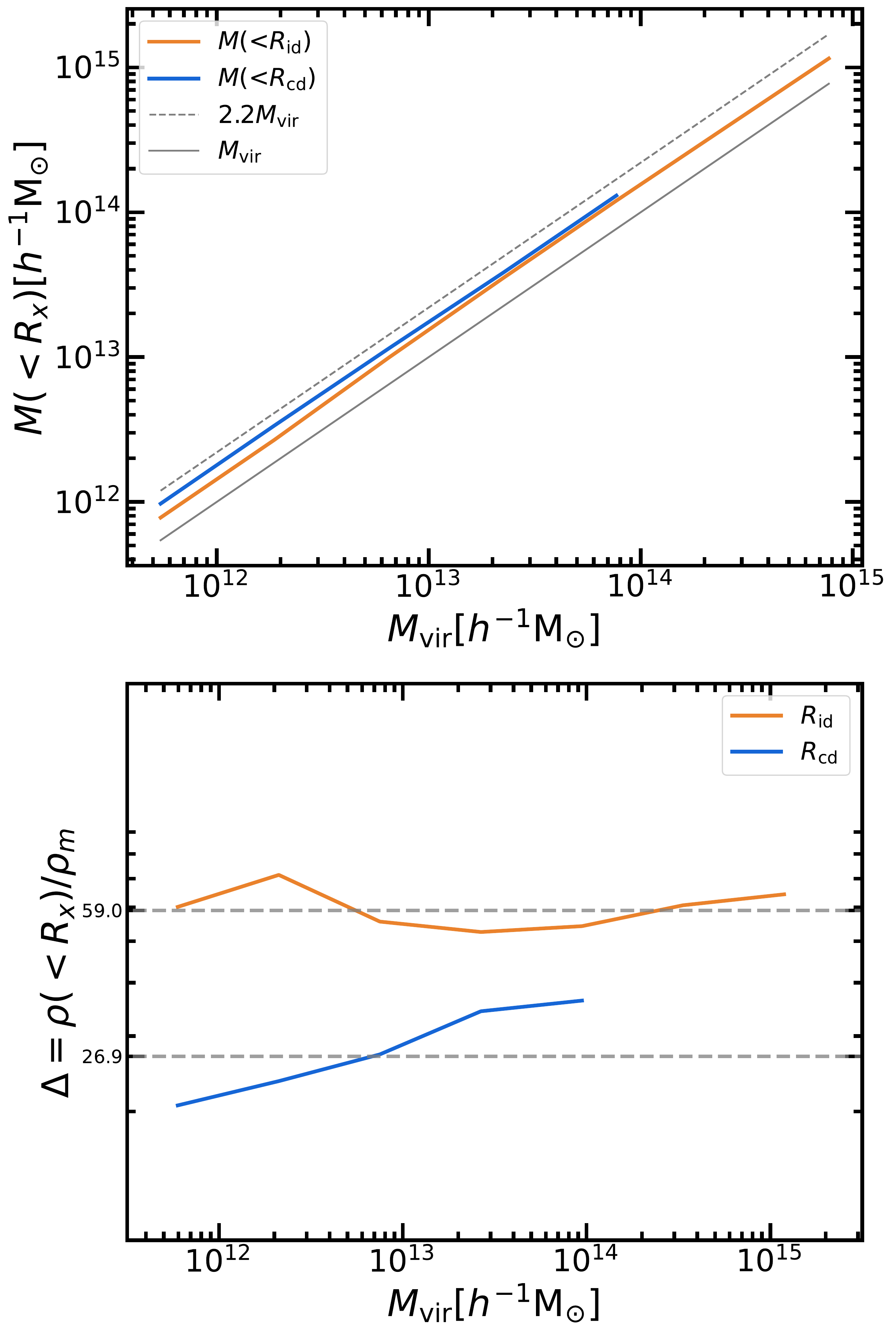}
    \caption{\textit{Top}: The mass enclosed in $R_{\rm id}$ and $R_{\rm cd}$ as functions of the virial mass, computed from our model profiles. Grey solid and dashed lines represent $M_{\rm vir}$ and $2.2M_{\rm vir}$ respectively. \textit{Bottom}: The average density enclosed in $R_{\rm id}$ and $R_{\rm cd}$ from the model profiles. Grey dashed lines represent the mean densities averaged over all mass bins. The mass and average density enclosed in $R_{\rm cd}$ is estimated in MB1 to MB5.} 
    \label{fig:rhorelation}
\end{figure}

In Figure~\ref{fig:rhorelation} we explore the integrated masses and densities of the model profiles out to the different characteristic radii for the depletion sample. The $R_{\rm id}$ encloses the mass of $\sim 1.5$ times the virial mass for all mass bins, while the ratio between the mass within the $R_{\rm cd}$ and the virial mass is more complex. The average density within $R_{\rm id}$ roughly keeps a constant of $\rho(< R_{\rm id})=59\rho_m$, but corresponding densities within $R_{\rm cd}$ varies with halo mass. Based on the FoF sample, \citet{fong2021natural} found that both $R_{\rm id}$ and $R_{\rm cd}$ correspond to constant enclosed densities. In this work, the average density within the $R_{\rm cd}$ is suppressed at the low mass end, due to the cleaning of the halo sample. %A qualitative explanation for the contradiction is due to the use of different halo samples. As the hard exclusion scheme expands the $R_{\rm cd}$ of the given mass bin, the average density within the $R_{\rm cd}$ becomes smaller, and this effect is particularly pronounced for the small mass end so that the average density of the small mass halo is positively correlated with the virial mass. \textbf{(The explanation for high mass end still being sought, maybe due to the different numerical mothed to find the minimum)}

\section{Discussions}
\label{sec:sec5}

\subsection{Impacts of the exclusion radius}
\label{sec:percolation}

We have built a halo model based on the $R_{\rm id}$ boundary in the previous sections. The distribution of haloes in the depletion sample exhibits clear exclusion features and can be well described by a power-law with a sharp truncation. Using the Einasto profile without truncation helps us to construct a self-consistent model to predict the bias profiles accurately. The success of our model can be attributed to two key points. First, the hard-sphere exclusion removes low-mass haloes that are stripped by large-mass neighbours and behave abnormally, making it possible to describe the halo-halo correlation function succinctly. Second, the Einasto profile without truncation turns out to match with the present halo sample and performs well in decomposing the density field, which is shown visually in Figure \ref{fig:Results1}. 

Although the current model with the $R_{\rm id}$ boundary is proven effective, it still leaves some issues worth discussing. For example, is the $R_{\rm id}$ exclusion criterion the only possible solution? Is the Einasto profile still appropriate for a self-consistent model with other exclusion criteria? A systematic investigation of these questions is not a trivial task. \citet{garcia2021redefinition} showed that by optimizing a multi-parameter model including the halo exclusion effect, an optimal halo exclusion radius is found near the minimum of $r^2\xi_{\rm hm}$. They also found that this optimal exclusion radius is about $1.3R_{\rm sp,87}$\footnote{$R_{\rm sp,87}$ is the radius containing 87\% of the splashback radii of individual particles in  a halo~\citep{diemer2017splashback}.}, according to which \citet{fong2021natural} inferred that it is also very close to $R_{\rm id}$. These results partly answer the questions we raised here and support $R_{\rm id}$ as an optimal solution for the halo exclusion radius. Below in this subsection, we will show how results vary if other radii, such as $R_{\rm vir}$, are chosen as the halo-halo exclusion radius in our model.

First, we clean the total sample with the new boundary $R_{\rm vir}$. The resulting new sample (hereafter the $R_{\rm vir}$-based sample) removes about $0.1\%$ of the haloes. The parameters for modelling $\xi_{\rm hh}$ are modified to match the different exclusion criterion. The exclusion scale $r_{\rm t}=R^1_{\rm vir}+R^2_{\rm vir}$ and halo biases are measured from the new auto-correlation functions. Parameters $r_{0}$ and $\gamma$ remain the best-fitting value from Section~\ref{sec:CFhh_fit} to compare with the power-law in the depletion sample.  

\begin{figure*}
    \includegraphics[width=\textwidth]{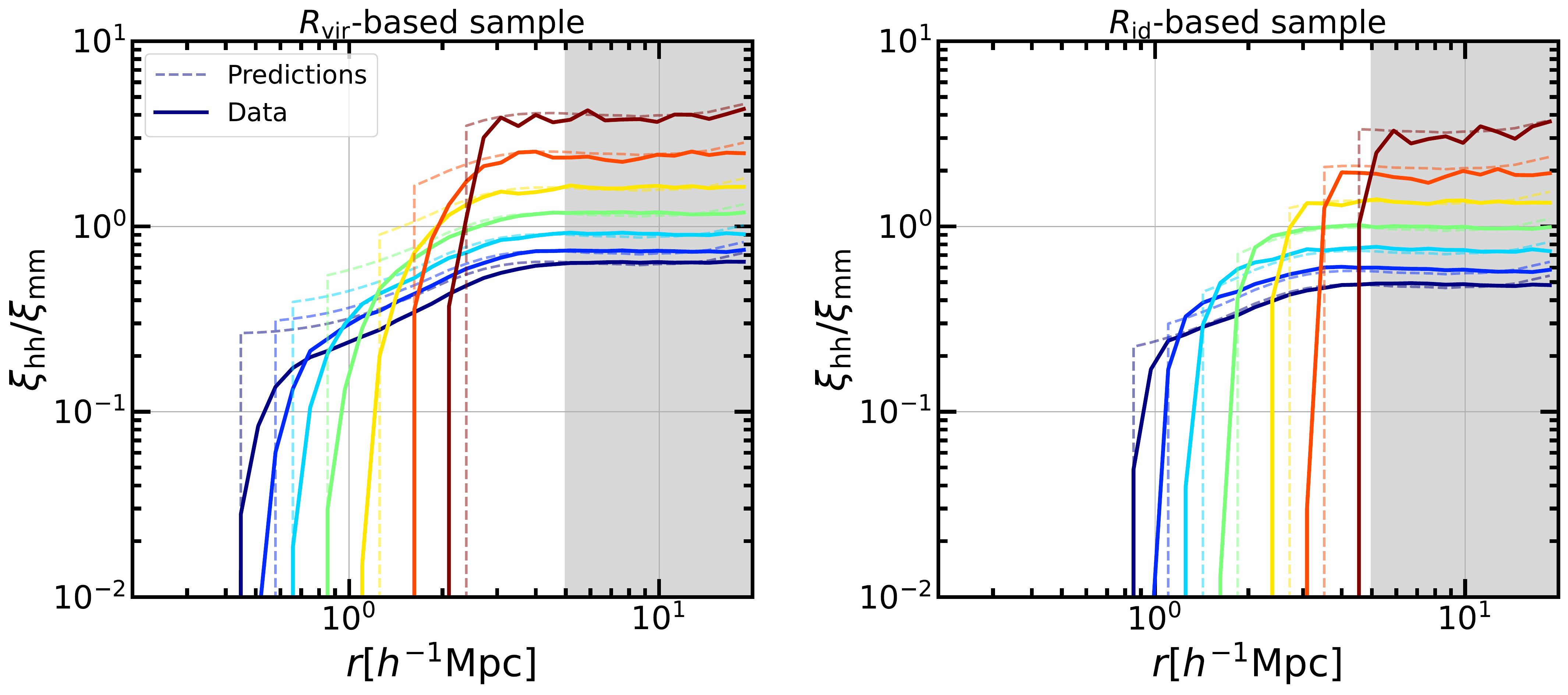}
    \caption{The halo-halo correlation in different exclusion catalogs. The y-axis represents the halo-halo correlation function scaled by the matter-matter correlation function. The left and right panel are the results of $R_{\rm vir}$-based sample and $R_{\rm id}$-based sample, respectively. The dashed lines are the predictions of our model. Grey regions marks the scales $r>r_0=4.96h^{-1}{\rm Mpc}$.}
    \label{fig:xihh_vir}
\end{figure*}

We measure the halo-halo correlation functions for the $R_{\rm vir}$-based sample and show the measurements of $\xi_{\rm hh}(m_1,m_2)$ and the predictions of the model in the left panel of Figure \ref{fig:xihh_vir}. For comparison, we show the results for the depletion sample ($R_{\rm id}$-based sample) and the corresponding predictions in the right panel. For $R_{\rm vir}$-based sample, at large scales, the model fits the measurements well, although the parameters $r_0$ and $\gamma$ are obtained from the data of the depletion sample. It suggests that the shape of $\xi_{\rm hh}$ is universal at large scales and is little affected by the definition of the halo exclusion boundary. Going to the trans-linear region ($r<5h^{-1}{\rm Mpc}$), the predictions of the model start to deviate from the measurements and the number densities of neighbouring haloes become lower than a power-law. %The locations of the cut-off measured in the virial sample may not perfectly match the characteristic scales of exponential decay\jx{you no longer have exponential decay in model. anyway, do they not match?}. For the high-mass bins, there is still a sharp truncation at the position of $R_{ex}$ despite the correlation function at the trans-linear region being suppressed\jx{what do you mean by being suppressed?}. For the lowest-mass bins, the cut-off position of the measurement is a little biased than the $R_{ex}$\jx{i fail to see this?}. 
%\jx{this may be because you did not isolate bridged haloes}
This extra suppression of the neighbour number density outside the exclusion scale implies the existence of some physical processes that affect the relative distribution of haloes as halo pairs become closer. For example, the tidal stripping can happen out to a distance of twice the virial radius of the central halo~\citep{behroozi2014mergers}, and many neighbouring haloes outside the virial radius are actually ejected subhaloes of the central host~\citep{Ludlow09}. The dynamics of haloes can become more complicated due to the interaction of their extended particle distributions, and dynamical friction also starts to set in as haloes become close. In addition, haloes can be bridged by the FoF algorithm when they are close to each other, which are regarded as merging ones in the catalog. All these processes can act to modify the spatial and mass distribution of neighbouring haloes, introducing additional complexity in the halo-halo correlation function near the virial scale. It might be possible to model the suppression at trans-linear scales by some more complicated functions, but how universal such modifications can be require further investigations.

%Such suppression does not appear to be due to different exclusion criteria. For the hard exclusion scheme changing the $R_{ex}$ can only change the position of the truncation and does not affect the distribution of haloes in the nearby region of $R_{ex}$. It is more likely that the halo inside of the $R_{\rm id}$ has an anomalous distribution of neighbours, causing the halo-halo correlation functions of the $R_{\rm vir}$-based sample has different patterns at intermediate scales and large scales.

To investigate the second question mentioned in this subsection, i.e. is the Einasto profile an appropriate profile for other boundaries, we refit the bias profiles in the $R_{\rm vir}$-based sample. The model framework and fitting procedure are the same as before but some parameters are modified. The halo biases and mass function are calculated using Equations \ref{eq:bias} and \ref{eq:MF}, but the best fitting values of the parameters are obtained from the data of the $R_{\rm vir}$-based sample. The truncation radius for $\xi_{\rm hh}$ changed for the new boundary definition while $r_0$ and $\gamma$ are the same as those used for the depletion sample. For the halo density profile, we consider two variants of the Einasto profile: the original Einasto profile without truncation, and the Einasto profile truncated at $R_{\rm vir}$. To avoid potential numerical problems, we use a sharp exponential decay to truncate the Einasto formula in 2-halo term to ensure the continuity of the profile. %the Einasto profile without truncation in both 1-halo terms and convolving of 2-halo terms.
%To answer the second question mentioned in this subsection, i.e. whether there is a self-consistent density profile with $R_{\rm vir}$ as the exclusion boundary, we refit the bias profile using our halo model based on $R_{\rm vir}$. We use the same model and fitting procedure, but modified some parameters. The halo biases and mass function are calculated using Equations \ref{eq:bias} and \ref{eq:MF}, but the best fitting values of the parameters are obtained from the data of the $R_{\rm vir}$-based samples. $R_{\rm ex}$ for $\xi_{\rm hh}$ becomes $R^1_{\rm vir}+R^2_{\rm vir}$ while $r_0$ and $\alpha$ are the same as those used for the depletion sample, since the shape of $\xi_{hh}$ is universal as shown earlier. For the halo density profile, we consider three variants of the Einasto profile: the original Einasto profile without truncation, and the Einasto profile multiplied by a decay function
%\begin{equation}
    %D(r)={\rm exp}\left[ -\left(\frac{r}{R_{\rm vir}}\right)^{d_{\rm decay}}\right],
%\end{equation}
%\jx{this equation is incorrect. what's the decay scale?}where decay factors $d_{\rm decay}$ are 2 and 4, respectively. \jx{what is the decay factor? specify the equation} 

\begin{figure*}
    \includegraphics[width=\textwidth]{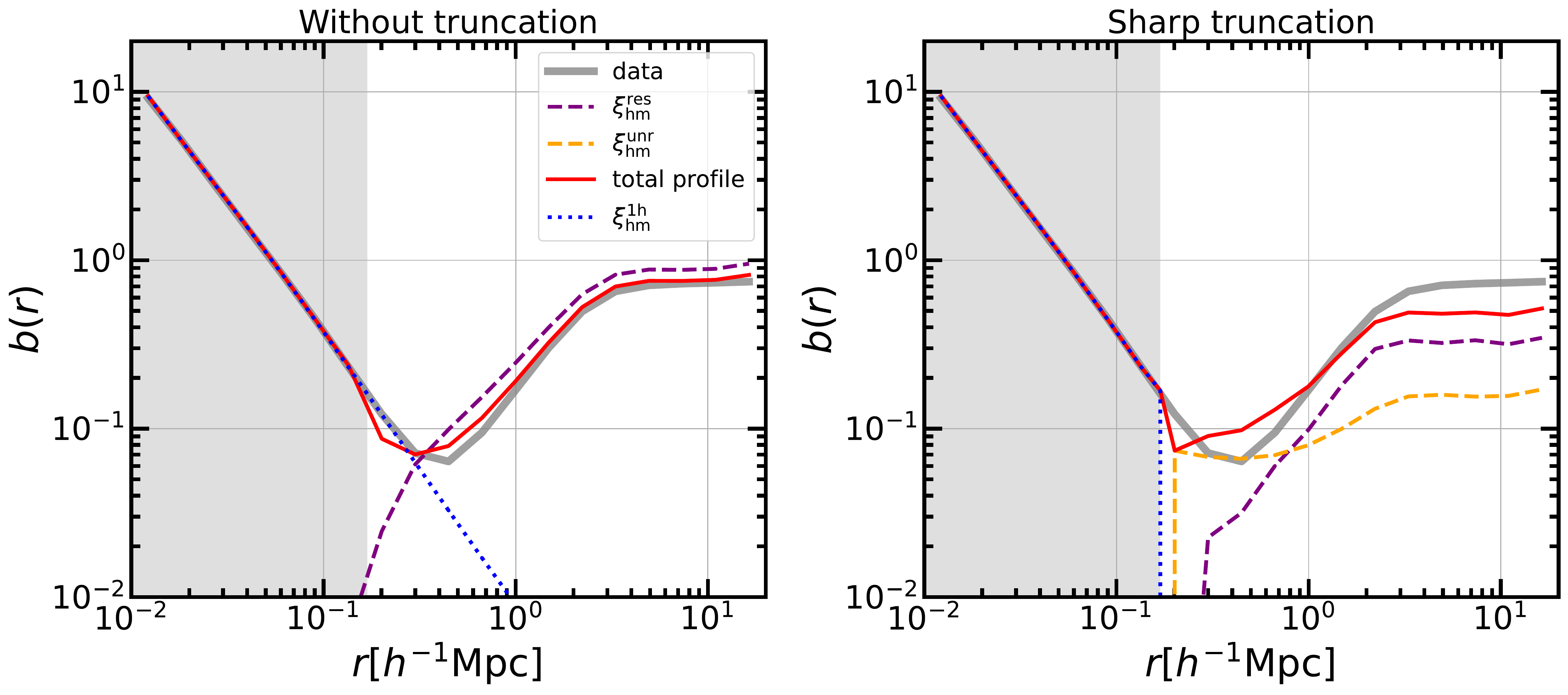}
    \caption{\textit{Left}: The halo bias profile using $R_{\rm vir}$ as the exclusion radius. %\jx{explain the vertical line}
    The grey and red solid lines represent the bias profiles of MB1 from data and model in the $R_{\rm vir}$-based sample. The purple and orange dashed lines represent the resolved and unresolved terms, respectively. The blue dotted line represents the 1-halo term. For the Einasto profile without truncation, $b_{\rm unr}$ is negative, so the unresolved terms does not appear in the left panel. The grey region marks the scales smaller than $R_{\rm vir}$.  %\jx{this plot suggests you need a softer truncation, and it may still be possible to find such a solution if you optimize your decay factor}
    \textit{Right}: same as the left except that the halo profile is the Einasto profile truncated at $R_{\rm vir}$ in the calculation of the 1-halo and resolved terms. }
    \label{fig:results_vir}
\end{figure*}

We fit the bias profile by optimizing the parameters $b_{\rm unr}$ in the $R_{\rm vir}$-based sample. The fitting results are shown in Figure \ref{fig:results_vir}. The bias profile can be well fit on both large and small scales in the $R_{\rm vir}$-based sample by adopting the Einasto profile without truncation. On the intermediate scale, some deviations between the fit and the data are observed, which are expected given the deviations seen in the halo-halo correlation function. Nevertheless, the main issue with this model is that the parameter $b_{\rm unr}$ is negative in the fit. It means the adopted density profile overestimates the total mass in the universe even when only counting the resolved haloes. This is not surprising because the virial catalog contains haloes that are much closer to the central one than in the depletion catalog. These haloes are counted in the outer part of the Einasto profile in the depletion catalog. Switching to the virial catalog, the Einasto profile has to be truncated or suppressed to avoid double-counting these masses. 

For the truncated Einasto profile, our model fails to predict the bias profile on both intermediate and large scales. Beyond the halo boundary, there is still a considerable fraction of mass distributed in these regions, which is not included in the truncated Einasto profile, so the clustering of matter on large scales is underestimated. The unresolved term can partly compensate for the missing matter at large scales by adjusting $b_{\rm unr}$, but at the expense of overpredicting the intermediate scale clustering. This contradiction illustrates that at least some of the matter outside $R_{\rm vir}$ is an integral part of the halo and cannot be modelled by diffuse matter.  Mathematical solutions may exist for the proper profile corresponding to a given boundary definition, although the solution might be complex or unphysical~\citep[e.g.,][]{2020PhRvD.101j3522C}. A complete answer to this question is beyond the scope of this paper, and we will advance such investigations in subsequent works. %There are many "subhaloes" in the region before $R_{\rm vir}$ and $R_{\rm id}$, as we have seen in the previous discussion. This part of the mass is double-counted by the 1-halo term and 2-halo term in the first scenario so that the bias profile outside $R_{\rm vir}$ is overestimated by the model. 

%For the exponentially decaying Einasto profiles, we find that even optimizing the amplitude of the unresolved matter, the final results underestimate the halo bias at linear scales and overestimate the bias trough in the trans-linear region. Increasing the contribution on the linear scale requires a more extended density profile to include more mass into the halo, which will lead to an overestimation in the trans-linear region, and vice versa. %This contradiction %shows that the trend of $\xi_{hh}$ is problematic because it is impossible to fit the bias profiles at intermediate and large scales at the same time by adjusting the Einasto profile, which implies that using $R_{\rm vir}$ as the exclusion radius may not be a suitable choice for a simpler, self-consistent halo model. 

In summary, although $R_{\rm id}$  may not be the only solution for halo exclusion in our model framework, there are some difficulties in constructing a self-consistent model using the traditional boundary, for example, $R_{\rm vir}$, as the exclusion radius. The cost of the $R_{\rm vir}$-based model is the need for a more complex description of the masses outside the $R_{\rm vir}$, as well as a more complex model for the halo-halo correlation function. We have also tried the $R_{\rm 200m}$ as the boundary definition, and the resulting conclusions are similar to that from the $R_{\rm vir}$-base sample. In contrast, the $R_{\rm id}$-based model involves only simple and natural forms of the density profile and $\xi_{\rm hh}$, and the fit performs the best among our experiments. This strengthens our expectation that the physical boundary found from the dynamical signature of halo accretion is also a good fit for partitioning mass into haloes.

\subsection{Interpreting the characteristic depletion radius from halo exclusion}
According to the dynamical features around the halo boundary, \citet{fong2021natural} interpreted $R_{\rm id}$ as the inner boundary of the depletion region, while $R_{\rm cd}$ is the location where the material is depleted the most. Turning to this work, more physical insights into the depletion radius can be dug out from the perspective of the halo model. %At large scales, we are concerned with how haloes interact with each other. When considering two haloes close to each other, their separation must be larger than a certain scale otherwise they merge. Since the hard-sphere scheme is adopted in our model, $R_{\rm id}$ can be interpreted as the halo-halo exclusion radius to prevent overlap. The reason for choosing Rid as the boundary is that we expect the objects in the halo catalog to have clear features at the scale of the depletion trough.

\begin{figure}
    \includegraphics[width=\columnwidth]{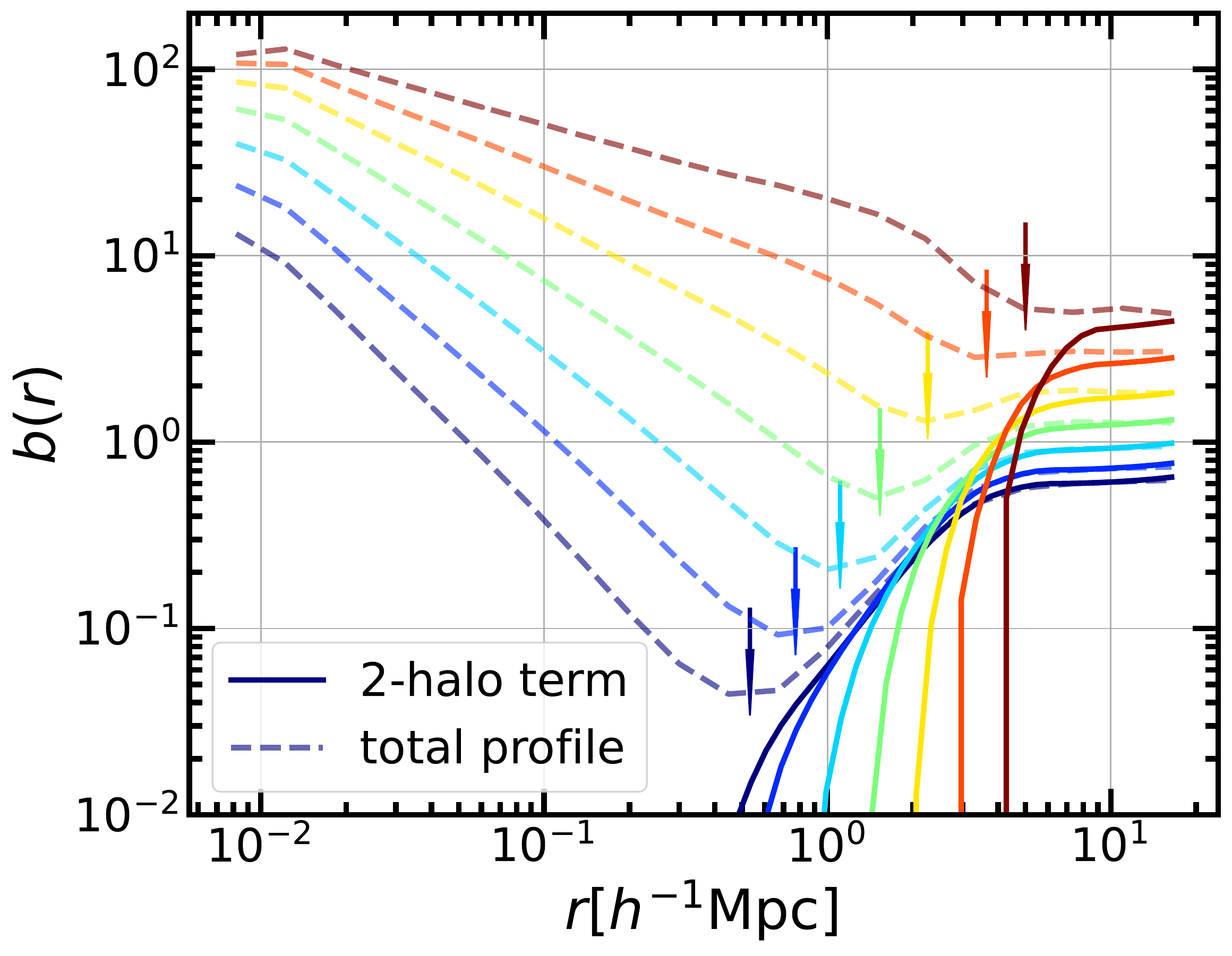}
    \caption{The 2-halo term contributions to the bias profiles around haloes in different mass bins. Solid coloured lines are the results from our model. Dashed lines are the measured total bias profiles. Arrows mark the locations of $R_{\rm cd}$'s.}
    \label{fig:2halo}
\end{figure}

In the halo model, the characteristic depletion radius defined at the bias minimum is located where the 1-halo term intersects with the 2-halo term. As low mass haloes form early, they are currently in a late phase of halo evolution with their density profiles extending well into a much depleted low density region. On the other hand, due to halo exclusion, the 2-halo term contributed by haloes with $m\gg M$ starts at a large distance relative to the radius of the central halo. This leads to a wide transition region in the 2-halo term around low mass haloes, which appears between the linear scale and the halo boundary as shown in Figure~\ref{fig:2halo}. A low outer density in the 1-halo term and a wide decay region in the 2-halo term together result in a clear trough in the bias profile. %For low mass bins, the 2-halo term has a transition phase between the halo boundary and linear scales, and the bias profile at these scales roughly follows the shape of the 2-halo term but is also affected by the central halo extension. 
As the mass increases, the transition interval in the 2-halo term becomes narrower and the contribution of the 1-halo term becomes significant, leading to a blurring of the depletion features. In extreme cases, $R_{\rm cd}$ becomes ill-defined when estimated from the total profile. Thus we use the typical value $2.5R_{\rm vir}$ \citep{fong2021natural} as the proxy of $R_{\rm cd}$ for MB6 and MB7. In Figure \ref{fig:2halo} we find that for all mass bins, the 2-halo term decreases rapidly near $R_{\rm cd}$, implying that it may be possible to extract $R_{\rm cd}$ from just the transition phase.

To this end, we develop a new phenomenological fitting formula to describe the inner and outer density profiles,
\begin{align}
    \label{eq:growth_term}
    \rho(r) &= \rho_{\rm EIN}(r) + \rho_{\rm outer}\times \Phi\\
    \rho_{\rm EIN}(r) &= \rho_s {\rm exp}\left( -\frac{2}{\alpha}\left[ \left(\frac{r}{r_s}\right)^\alpha-1 \right]\right)\\
    \rho_{\rm outer}(r) &= [\rho_1\left( \frac{r}{r_1} \right) ^ {-\gamma_1}+\rho_2] \\
    \Phi(r) &={\rm exp}\left[ -\frac{1}{(r/r_{\rm decay})^4} \right]
\end{align}
This analytic formula describes the inner halo profile by the standard Einasto profile. The outer part of the halo profile, $\rho_{\rm outer}$, is described by a constant density $\rho_2$, plus a power law. %The parameters $\rho_1$ and $\eta$ characterize the normalization and index of the power law, respectively. 
Note the pivot radius $r_1$ and $\rho_1$ are degenerate, so that one of them can be fixed arbitrarily. %and the fitting results are not sensitive to the choice of the pivot radius, so that we artificially choice $R_{\rm cd}$ as the pivot radius of power law.
%\jx{the Rcd as pivot is not a good choice if you want to use the fitting formula to extract Rcd.} 
The decay term $\Phi(r)$ is described by an exponential decay. The parameter $r_{\rm decay}$ is the characteristic radius of the decay term. 

\begin{figure}
    \includegraphics[width=\columnwidth]{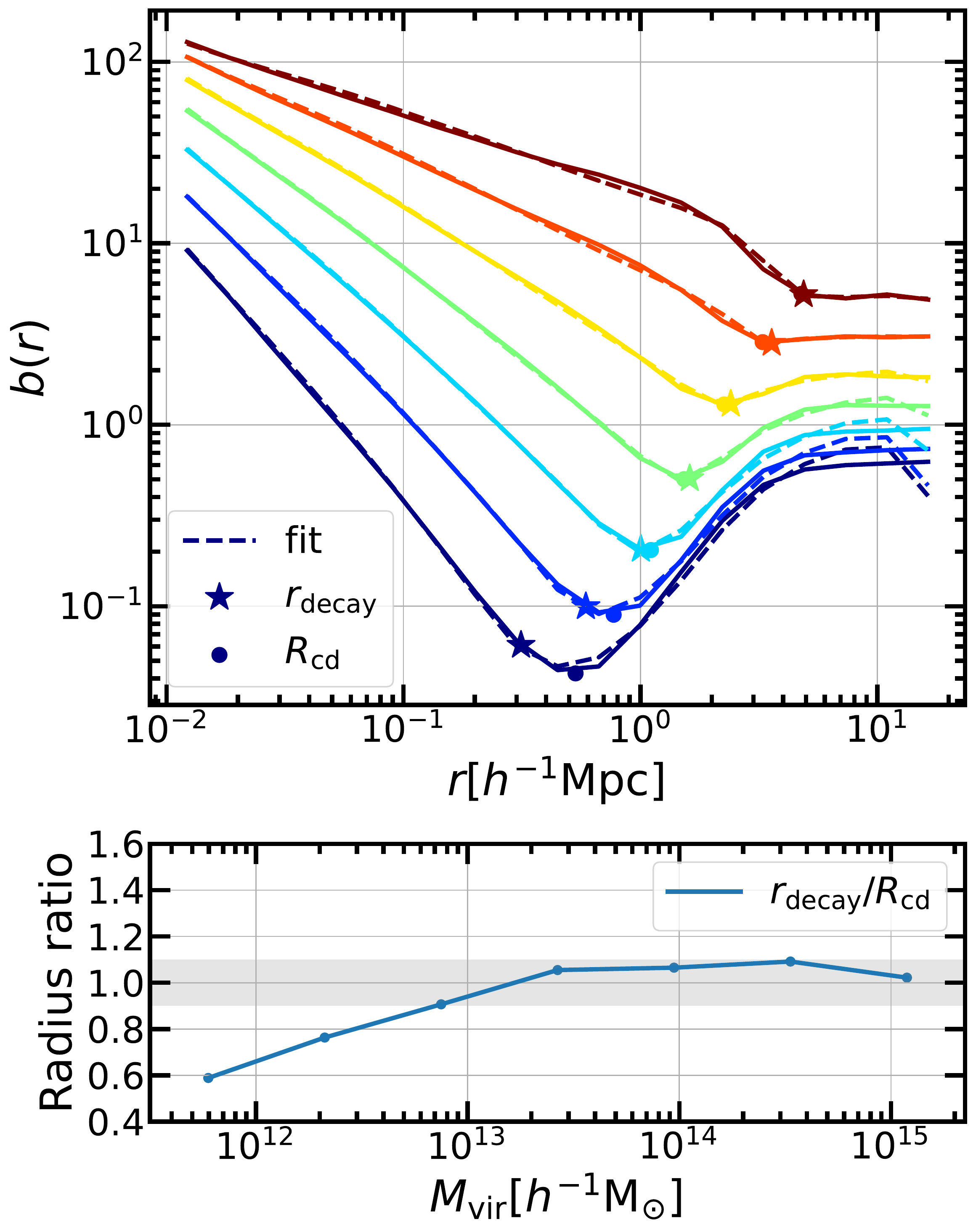}
    \caption{\textit{Top}: fitting results of equation \ref{eq:growth_term}. Stars and Dots mark the location of $r_{\rm decay}$ and $R_{\rm cd}$ ($2.5R_{\rm vir}$ for MB6 and MB7), respectively. \textit{Bottom}: ratio between $r_{\rm decay}$ and $R_{\rm cd}$ in different mass bins. Grey region marks the $1\pm 0.1$ range.}
    \label{fig:fitting}
\end{figure}

In Figure \ref{fig:fitting} we show the fitting results of equation \ref{eq:growth_term}. We restrict $r_{\rm decay}$ in the range $0.5- 5R_{\rm id}$ and convert the results to the bias profile. We see that our fitting formula provides a good description of the simulation data, except for the two lowest mass bins. For low mass haloes, the power law does not describe the outermost profile well, similar to the results of \citet{diemer2014dependence} in Figure \ref{fig:result_comp} which has a similar form in the outer profile. However, the depletion trough features (the minimum of bias and its location) are well preserved. We also label the locations of the $R_{\rm cd}$ ($2.5R_{\rm vir}$ for MB6 and MB7) and characteristic radius. Their relative deviation is less than 0.1 for $M_{\rm vir}\gtrsim 10^{13}\msunh$, as shown on the bottom panel of Figure \ref{fig:fitting}. This indicates that $r_{\rm decay}$ can be used as a good proxy for $R_{\rm cd}$ for medium and large mass haloes. This conclusion can help us to estimate the $R_{\rm cd}$ of supermassive haloes for which the depletion feature is not clear due to their early growth phase.

%In summary, $R_{\rm id}$ is artificially set to the halo-halo exclusion radius in our model, describing the truncation of the neighbouring halo distribution. The contributions of haloes with different mass are accumulated to form the 2-halo term, which retains the exponential truncation feature. $R_{\rm cd}$ characteristic the truncation of the total 2-halo term. When adopting equation \ref{eq:growth_term} to fit the halo density profile, $R_{\rm cd}$ can be considered as the effective truncation radius of the 2-halo term. Combined with the analysis of the minimum of the bias profile, our method allows for a more accurate estimate of the $R_{\rm cd}$ for high mass haloes. 
In contrast to DK14 or other similar models, our model does not suppress the 1-halo term, so the transition terms of these models are interpreted differently from the decay terms of our model. The transition term in DK14 describes the behavior of the orbital component at the edge, while our decay term describes the behavior of the 2-halo term near the halo boundary. The different division of the halo outer profile reflects different perspectives on the definition of a halo. We hope to consider the halo and its active growth region as an inseparable whole, which facilitates a better understanding of halo growth, as well as a concise description of the large-scale structure of the universe.

\section{Summary and conclusions}
\label{sec:sec6}
In this work we build a halo model for the halo-matter correlation function based on the inner depletion radius as a physical halo boundary. 

The inner depletion radius separates the growing part of a halo from a decaying environment, and is expected to be an optimal choice for modelling halo exclusion~\citep{fong2021natural}. This radius is defined where the mass infall rate peaks, and is typically found at $\sim 2$ times the virial radius. 

Starting from the FoF halo catalog in a high resolution $N$-body simulation, we first build a depletion catalog by removing haloes that overlap with more massive neighbours. 
%selecting only haloes that do not overlap out to their inner depletion radius. 
We then extract various halo-model components based on this depletion catalog, including the halo density profile, halo-halo correlation function, halo bias and mass function.
Putting these components together we obtain a self-consistent halo model that accurately describes the halo-matter correlation function across the linear and non-linear scales, for haloes in the mass range of $10^{11.5}\msunh<M_{\rm vir}<10^{15.35}\msunh$. 
In particular, halo exclusion is explicitly accounted for in the halo-halo correlation function which excludes haloes from overlapping on their depletion radii. Convolving the halo-halo correlation with the halo density profile, the matter distribution around the transition scale is accurately recovered, with no need to appeal to additional radial-dependent bias or exclusion probability function as in previous works. The main recipes and conclusions of our model are the following.

\begin{itemize}
    \item Adopting the inner depletion radius as a halo boundary, the density profile of each halo is described by the Einasto profile (Equation~\eqref{eq:EIN}). Within the halo boundary, we show that the Einasto profile can well fit the density profile from the simulation. The analytical profile extends beyond the boundary. This extended outer profile is needed to account for the merging mass associated with the halo, including those removed from the FoF catalog due to their overlapping boundaries with others. The good performance of our model suggests that the Einasto profile is a natural dual to the $R_{\rm id}$-based halo exclusion.
    \item The halo-halo correlation function in the depletion catalog follows a universal shape, which can be parametrized by a unit halo-halo correlation function multiplied with mass-dependent but scale-independent biases. The unit halo-halo correlation function, $\hat{\xi}_{\rm hh}$, is a simple power law (Equation~\eqref{eq:xi_hh_star}), and describes the correlation function for haloes with a unit bias, $b(m_\ast)=1$. Each halo-halo correlation function is truncated at a truncation radius equalling the sum of the depletion radii of the central and neighbouring haloes, reflecting the effect of halo exclusion in our depletion catalog.
    \item The universal halo-halo correlation also enables us to model the contributions from unresolved haloes as well as from diffuse matter to the density field. Adding their contributions together results in an unresolved term in the halo-matter correlation function (Equation~\eqref{eq:fit_unr2}), whose shape follows the unit correlation function. Because the unresolved haloes are much smaller compared to the central halo that we try to model, they can be treated as point masses together with diffuse matter. This leads to a truncation radius in the unresolved term right at the depletion radius of the central. The effective bias of the unresolved term is left as a free parameter in our model. However, this parameter can also be estimated from mass conservation if we ignore the effect of second-order bias.
    \item The 2-halo term is dominated by the contributions from group to cluster mass haloes on the linear scale. The contributions from low mass haloes are important in shaping the transition region near the exclusion radius. The unresolved haloes and diffuse matter only have a percent level contribution to the correlation function across all scales.
    \item The halo bias required to accurately model the halo-halo correlation function involves contribution from high-order biases. We extract this bias function (Equation~\eqref{eq:bias}) from the halo-halo correlation measurements on large scale and fit it using the fitting formula of \citet{jing1998accurate}.
    \item The halo mass function in our depletion sample differs from that in the FoF sample significantly on the low mass end. Despite this, both can be well fit using the Sheth-Tormen~\citep{sheth1999large} formula. 
    \item The final model can fit the halo-matter correlation function with an accuracy of $< 9 \%$ across the linear and highly non-linear scales. In particular, it can well describe the transition in correlation function around the halo boundary, in contrast to some previous models that have difficulties in accurately modelling this region. 
    \item The resulting model profile can also be used to accurately predict the locations of other characteristic radii in the halo profile, including the splashback radius and the characteristic depletion radius. Both radii can be predicted to an accuracy of $\lesssim 10\%$. Our model also provides more insights into the characteristic radius at the bias minimum, which exists as a gluing radius for the 1-halo and 2-halo terms. Around high mass haloes where the bias minimum is not well defined, this characteristic radius can be extracted from the decay scale of the 2-halo term alone.
\end{itemize}

While we have adopted the depletion radius as a physically-motivated radius for halo exclusion, our current work does not exclude alternative choices for halo exclusion that may perform similarly well. However, as we have shown, trying to use the virial radius for halo exclusion meets the difficulty of simultaneously matching the small scale and large scale clustering. Coupling the virial radius with a non-truncated Einasto profile leads to an overflow of mass on large scale, violating the self-consistency of the model. Such an overflow is not realized in conventional models because mass conservation is not explicitly modelled in the classical model. In addition, the halo-halo correlation function also takes on a more complicated form when using the virial radius to dissect haloes, potentially due to the stronger interactions between haloes on the virial scale. Moreover, we note that \citet{garcia2021redefinition} showed the existence of an optimal halo exclusion radius, which happens to be almost identical to the inner depletion radius we are using~\citep{fong2021natural}. In future work, we plan to investigate the uniqueness of the halo exclusion solution in a mathematical framework.

To facilitate comparisons with previous works and to serve as a transition from the classical halo definition, in the current model we still label each halo using the virial mass. Note our parametrization allows for any arbitrary mass definition as a tag to distinguish different haloes, as long as the corresponding halo boundary and density profile can be connected to this mass. With more understandings and developments on the depletion radius based halo, in the future we expect to eventually switch to the depletion mass as a halo label, to obtain a more self-contained halo model under the new halo definition.

\section*{Acknowledgements}
We thank Y.P. Jing for access to the CosmicGrowth simulation. YF thanks Hongyu Gao for help with the calculation of the velocity profile. This work is supported by NSFC (11973032, 11890691, 11621303), National Key Basic Research and Development Program of China (No.\ 2018YFA0404504), 111 project (No.\ B20019), and the science research grants from the China Manned Space Project (No.\ CMS-CSST-2021-A03). We thank the sponsorship from Yangyang Development Fund. The computation of this work is done on the \textsc{Gravity} supercomputer at the Department of Astronomy, Shanghai Jiao Tong University. 

%%%%%%%%%%%%%%%%%%%%%%%%%%%%%%%%%%%%%%%%%%%%%%%%%%
\section*{Data Availability}
The data underlying this article will be shared on a reasonable request to the corresponding author.


\bibliographystyle{mnras}
\bibliography{example} % if your bibtex file is called example.bib

%\begin{thebibliography}{99}
%\end{thebibliography}

%%%%%%%%%%%%%%%%%%%%%%%%%%%%%%%%%%%%%%%%%%%%%%%%%%

%%%%%%%%%%%%%%%%% APPENDICES %%%%%%%%%%%%%%%%%%%%%

\appendix

\section{Properties of excluded sample}
\label{app:appA}
In this appendix, we show more differences between the depletion sample and the excluded sample. We focus on a list of properties describing the halo structure from different aspects, they are

\begin{itemize}
    \item $V_{\rm max}/V_{\rm vir}$: The maximum of the circular velocity function;
    \item $j$: The spin of the central subhalo;
    \item $e$: The shape parameter of the halo;
    \item $a_{1/2}$: The formation time paramter;
    \item $\delta_{\rm e}$: The environment overdensity of the halo;
\end{itemize}

The specific definitions of these halo properties are given in \citet{han2019multidimensional}. In this appendix, we will explore their dependences on mass in the different samples, which helps us to better understand the halo exclusion.

\begin{figure*}
    \includegraphics[width=\textwidth]{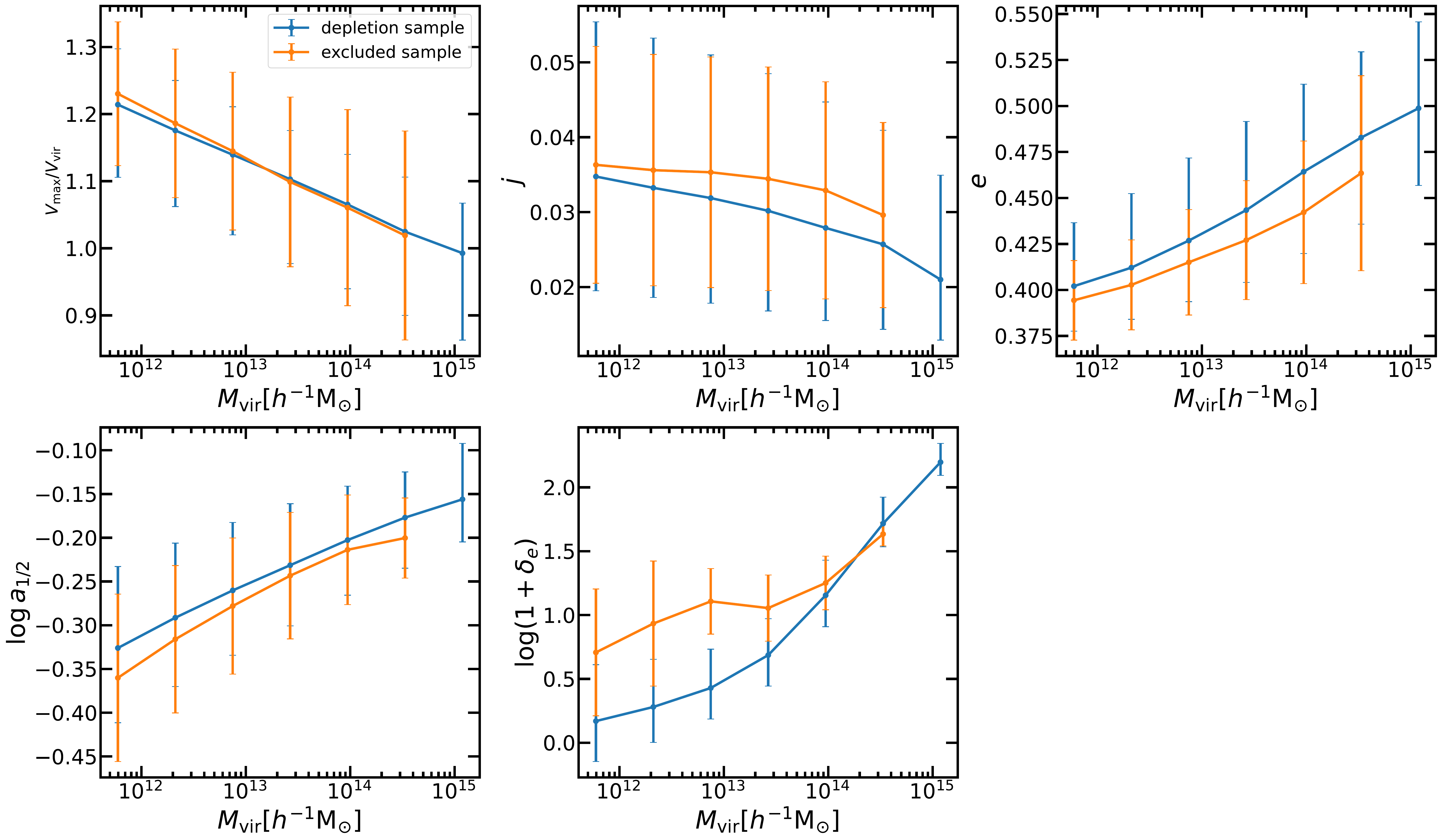}
    \caption{The mass dependence of halo properties in different halo samples. The data points are the medians of different mass bins. Error bars mark the 16-84th percentile range. For the excluded sample, there is no data point for MB7 since no halo is excluded from MB7 of FoF sample.}
    \label{fig:apdxA_1}
\end{figure*}

Figure \ref{fig:apdxA_1} shows the mass dependence of halo properties in different samples. Since no halo is excluded from the MB7 of the FoF sample, the excluded sample has no data points for MB7. For the parameters $V_{\rm max}/V_{\rm vir}$, there is no significant difference between the depletion and excluded samples. Concentrating on the median of the parameters, $e$ and $a_{1/2}$ in the excluded sample are lower than those in the depletion sample across different mass ranges, while $j$ in the excluded sample are higher than that in the depletion sample. %When focusing on the 16-84 percentile ranges of these three parameters, two samples are not well distinguished. 
Examining the environmental parameters of the different samples, we find that the low-mass haloes in the excluded sample have a larger environmental density, while for the high-mass halo, the environmental densities of the two samples are close to each other. The depletion and excluded samples are distinguished better in the environmental parameter space than in other spaces. 

\section{High order bias}
\label{app:hobias}
According to \citet{cooray2002halo}, the halo density field can be expanded as a function of the matter field,
\begin{equation}
    \delta_{\rm h}(m)=\sum_i b_i(m)\delta^i.
\end{equation}
The conventional treatment on large scales is to only consider the linear bias, but this approximation will not hold when we compare $b_{\rm hm}$ and $b_{\rm hm}$. Considering second order bias $b_2$, the auto-halo correlation can be written as,
\begin{align}
    \langle \delta_{\rm h}\delta_{\rm h}\rangle & =  \langle(b_1\delta+b_2\delta^2)(b_1\delta'+b_2\delta'^2)\rangle \nonumber\\
    & =  b_1^2\langle \delta\delta'\rangle+b_1b_2\langle \delta\delta'^2\rangle\nonumber\\
    & +b_1b_2\langle \delta^2\delta'\rangle+b_2^2\langle \delta^2\delta'^2\rangle .
    \label{eq:highhh}
\end{align}
Similarly, the halo-matter correlation is rewritten as,
\begin{equation}
    \langle \delta_{\rm h}\delta_{\rm m}\rangle=b_1\langle \delta\delta'\rangle+b_2\langle \delta^2\delta'\rangle
    \label{eq:highhm}
\end{equation}
Equations \ref{eq:highhh} and \ref{eq:highhm} clearly shows that $\langle \delta_{\rm h}\delta_{\rm h}\rangle$ and $\langle \delta_{\rm h}\delta\rangle$ are not only determined by the 2-point correlation and the linear bias but also affected by the high-order correlations and biases. As a result, the linear bias estimators $b_{\rm hm}$ and $b_{\rm hh}$ are not equivalent when considering $b_2$ at large scales. To quantify the deviations between $b_{\rm hm}$ and $b_{\rm hh}$, we simplify equations \ref{eq:highhh} and \ref{eq:highhm} by assuming the overdensity field is Gaussian at sufficiently large scales. Using the Wick theorem, we have
\begin{align}
    \langle\delta^2\rangle &= \langle\delta'^2\rangle = \sigma^2 \nonumber \\
    \langle\delta^2\delta'\rangle&=\langle\delta\delta'^2\rangle=0 \nonumber \\
    \langle\delta^2\delta'^2\rangle&=\sigma^4+2\langle\delta \delta'\rangle^2,
\end{align}
So that we obtain,
\begin{align}
    \label{eq:res1}
    \langle\delta_{\rm h}\delta\rangle&=b_1\langle\delta\delta'\rangle\\
    \langle\delta_{\rm h}\delta_{\rm h}\rangle&=b_1^2\langle\delta\delta'\rangle+b_2^2(\langle\sigma^4\rangle^2+2\langle\delta\delta'\rangle^2)
    \label{eq:res2}
\end{align}
For the cross-halo correlation, we have
\begin{align}
    \langle\delta_{\rm h}(m_1)\delta_{\rm h}(m_2)\rangle=&b_1(m_1)b_1(m_2)\langle\delta\delta'\rangle \\
    &+b_2(m_1)b_2(m_2)(\sigma^4+2\langle\delta\delta'\rangle^2).
\end{align}
These results indicate that the value of $b_{\rm hm}$ responds to $b_1$, while the value of $b_{\rm hh}$ responds to both of $b_1$ and $b_2$.

\section{Mass conservation at large scales}
\label{app:mass_conser}

On sufficiently large scales where the bias is linear and the haloes can be approximated as point masses, the decomposition of a perturbed density into haloes can be written as
\begin{equation}
    \rho=\int \bar{\rho}f(m) [1+b(m)\delta] \ud m,
\end{equation} which reduces to the local mass conservation relation, Equation~\eqref{eq:local_conser}. However, the bias we use in modelling the halo-halo correlation is not exactly the linear bias, but $b_{\rm hh}$ which can differ from the linear bias up to a second-order bias term, 
\begin{equation}
b_{\rm hh}(m)=b(m)+\epsilon(m).
\end{equation}
Plugging the above relation into Equation~\eqref{eq:A_theo_exact}, the mass conservation in presence of the un-resolved term becomes

\begin{equation}
    1=\int^{m_{\rm max}}_{m_{\rm res}}f(m)b_{\rm hh}(m)\ud m+b_{\rm unr}+\int \epsilon(m)f(m)\ud m,
    \label{eq:Mass_cons}
\end{equation}
which is a more accurate version of Equation~\eqref{eq:A_theo}.

In contrast to the mass conservation in classical models, Equation~\eqref{eq:Mass_cons} has some additional details. For the first integral on the right hand side, the lower mass limit is $m_{\rm res}$ instead of 0, so the integral value is some number between 0 and unity, representing the density fraction of resolved haloes. In addition, the contribution of unresolved matter is also taken into account in our model, reflected in the second term $b_{\rm unr}$. Lastly, a non-linear bias term, $\epsilon(m)$, arises due to the use of $b_{\rm hh}$ in our model.

Some previous works \citep{2016PhRvD..93f3512S, 2020A&A...641A.130M, 2021MNRAS.502.1401M} derived a form very similar to Equation~\eqref{eq:Mass_cons}, although their derivation is based on a completely different assumption. They assume that all remaining mass, except that bounded in resolved haloes, is contained in haloes of mass exactly $M_{\rm min}$, and accordingly design a mass function to enforce mass conservation. In contrast, our assumption that the halo follows the same normalized spatial distribution whether they are below $m_{\rm res}$ or not seems more natural, which does not require artificial modifications to the mass function and bias.

If $\epsilon(m)$ is known, then $b_{\rm unr}$ can be solved from Equation~\eqref{eq:Mass_cons} under mass conservation. Without knowing $\epsilon(m)$, in our current model we choose to fit $b_{\rm unr}$ as a free parameter in all mass bins. Moreover, the mass fraction $f(m)$ as defined in Equation~\eqref{eq:mass_frac} is obtained by weighting haloes with the integrated mass, $M_{\rm int}=\int \rho_{\rm h}(\textbf{\textit{r}}|m) \ud^3 r$, instead of the virial mass, $m$. Since we do not truncate the density profile at the halo boundary, $M_{\rm int}$ differs from both the virial mass and the enclosed mass within $R_{\rm id}$. Note that the extension of the halo profile beyond our depletion radius has not been directly verified with simulation data in our current analysis, and the optimal 1-halo profile may differ from the Einasto form in the outer part. Such a potential difference leads to some extra uncertainty in the integrated mass, $M_{\rm int}$, and consequently an uncertainty in the $f(m)$ function of the mass conservation relation. This uncertainty further requires that we fit for $b_{\rm unr}$ from the data, instead of solving it from Equation~\eqref{eq:Mass_cons}.

Because the first and third terms on the right hand side of Equation~\eqref{eq:Mass_cons} are constants, mass conservation then dictates $b_{\rm unr}$ is uncorrelated with halo mass $M$. %In other words, $b_{\rm unr}$ can be expected to be a constant independent of the halo mass $M$ under the premise of mass conservation. 
This conclusion is essential for the fit of the halo-matter correlation function.% To achieve a higher precision, we leave $b_{\rm unr}$ as a free parameter and optimize it in all mass bins. Therefore, the best-fitting B is mass-independent, and mass conservation is automatically maintained in such a global fit. %To verify this inference, we develop a global fit for $A(M)/b_{\rm hh}(M)$, optimizing it by considering the bias profiles of all mass bins. Figure \ref{fig:mass_conserv} shows the results of the global fit, where the best-fitting value of $A(M)/b_{\rm hh}(M)$ is 0.016. Although mass conservation is enforced, our model performs up to the percentage level, except for some radial bins at large scales. This remarkable result indicates that our model is compatible with mass conservation, although we do not directly obtain a theoretical solution for $A(M)/b_{\rm hh}(M)$.
%\begin{figure}
    %\includegraphics[width=\columnwidth]{figures/result_fixA.pdf}
    %\caption{\textit{Top}: The bias profiles and the global fit. The solid lines are the measured profiles. The dashed lines is the fits obtained by setting $A(M)/b_{\rm hh}$ as a unknown constant for different halo mass and optimizing it in all bins. The best-fitting value is $A(M)/b_{\rm hh}=0.016$ . \textit{Bottom}: The relative errors between the data and the fits} 
    %\label{fig:mass_conserv}
%\end{figure}

%%%%%%%%%%%%%%%%%%%%%%%%%%%%%%%%%%%%%%%%%%%%%%%%%%

% Don't change these lines
\bsp	% typesetting comment
\label{lastpage}
\end{document}